\documentclass[twocolumn,showpacs,amsfonts,aps,prc,floatfix,%
superscriptaddress,nofootinbib]{revtex4-1}
\usepackage{amsmath}
\usepackage{bm}
\usepackage{graphicx}

\newcommand{\fm}{\mbox{fm}}
\newcommand{\Tdec}{T_\mathrm{dec}}
\newcommand{\Tc}{T_\mathrm{c}}
\newcommand{\Tchem}{T_\mathrm{chem}}
\newcommand{\Tsw}{T_\mathrm{sw}}
\newcommand{\VISH}{{\tt VISH2{+}1}}
\newcommand{\VC}{{\tt VISHNU}}
\newcommand{\U}{{\tt UrQMD}}
\newcommand{\con}{{\tt H$_2$O}}

\newcommand{\dNdy}{dN_\mathrm{ch}/dy}
\newcommand{\Npart}{N_\mathrm{part}}

\newcommand{\be}[1]{\begin{equation}\label{#1}}
\newcommand{\ee}{\end{equation}}

\newcommand{\eq}{{\,=\,}}

\usepackage{color}

\begin{document}

%%%%%%%%%%%%%%%%%%%%%%%%Front Matter%%%%%%%%%%%%%%%%%%%%%%%%%%%%%%%%%%
%%%%%%%%%%%%%%%%%%%%%%%%%%%%%%%%%%%%%%%%%%%%%%%%%%%%%%%%%%%%%%%%%%%%%%

\title{Viscous QCD matter in a hybrid hydrodynamic+Boltzmann approach}
\date{\today}

\author{Huichao Song}
\email[Correspond to\ ]{HSong@lbl.gov}

\affiliation{Nuclear Science Division,
             %MS 70R0319,
             Lawrence Berkeley National Laboratory, Berkeley,
             California 94720, USA}

\affiliation{Department of Physics, The Ohio State University,
             %191 West Woodruff Avenue,
             Columbus, Ohio 43210, USA}

\author{Steffen A. Bass}
\affiliation{Department of Physics, Duke University, Durham,
             North Carolina 27708, USA}

\author{Ulrich Heinz}
\affiliation{Department of Physics, The Ohio State University,
             %191 West Woodruff Avenue,
             Columbus, Ohio 43210, USA}

\begin{abstract}
A hybrid transport approach for the bulk evolution of viscous QCD matter
produced in ultra-relativistic heavy-ion collisions is presented. The
expansion of the dense deconfined phase of the reaction is modeled with
viscous hydrodynamics while the dilute late hadron gas stage is described
microscopically by the Boltzmann equation. The advantages of such a hybrid
approach lie in the improved capability of handling large dissipative
corrections in the late dilute phase of the reaction, including a realistic
treatment of the non-equilibrium hadronic chemistry and kinetic freeze-out.
By varying the switching temperature at which the hydrodynamic output is
converted to particles for further propagation with the Boltzmann cascade
we test the ability of the macroscopic hydrodynamic approach to emulate
the microscopic evolution during the hadronic stage and extract the
temperature dependence of the effective shear viscosity of the hadron
resonance gas produced in the collision. We find that the extracted values
depend on the prior hydrodynamic history and hence do not represent
fundamental transport properties of the hadron resonance gas. We conclude
that viscous fluid dynamics does not provide a faithful description of
hadron resonance gas dynamics with predictive power, and that both
components of the hybrid approach are needed for a quantitative description
of the fireball expansion and its freeze-out.
\end{abstract}
\pacs{25.75.-q, 12.38.Mh, 25.75.Ld, 24.10.Nz}

\maketitle

%%%%%%%%%%%%%%%%%%%%%%%%%%%%%%%%%%%%%%%%%%%%%%%%%%%%%%%%%%%%%%%%%%%%%%%%%%%%
\section{Introduction}
\label{sec1}
%%%%%%%%%%%%%%%%%%%%%%%%%%%%%%%%%%%%%%%%%%%%%%%%%%%%%%%%%%%%%%%%%%%%%%%%%%%%

The discovery that ultrarelativistic heavy-ion collisions at the
Relativistic Heavy-Ion Collider (RHIC) produce dense, color deconfined
matter that thermalizes quickly into a quark-gluon plasma (QGP)
\cite{Heinz:2001xi} and subsequently evolves like an almost perfect
liquid \cite{Kolb:2003dz,Gyulassy:2004vg,Gyulassy:2004zy,Arsene:2004fa}
with near-minimal viscosity \cite{Danielewicz:1984ww,Policastro:2001yc}
has generated intense interest in a quantitative determination of the QGP
transport properties \cite{Teaney:2003kp,Romatschke:2007mq,Song:2007fn,%
Song:2008hj,Song:2010mg}. Since the QGP liquid stage is sandwiched between
an early pre-equilibrium and a final hadronic rescattering and decoupling
stage, both of which have different transport properties, such a
determination requires a complete dynamical description of all stages of
the fireball expansion \cite{Song:2008hj}. Stated differently, when one
uses experimental observables that are sensitive to the transport
properties of the expanding medium (for example, the elliptic flow $v_2$
has been shown to be particularly sensitive to shear viscosity
\cite{Teaney:2003kp,Romatschke:2007mq,Song:2007fn,Song:2008hj,Song:2010mg,%
Molnar:2001ux}), contributions from the pre-equilibrium and hadronic
freeze-out stages to these observables must be accurately known for a
precise determination of the QGP transport coefficients. Purely hydrodynamic
calculations, that substitute hydrodynamic initial conditions for a full
dynamical solution of the pre-equilibrium stage and replace the kinetic
hadron freeze-out process by a sudden transition from thermalized fluid
to non-interacting particles using the Cooper-Frye prescription
\cite{Cooper:1974mv}, require additional parameters whose values
may influence the extracted thermal and transport properties, and they
will always be plagued by irreducible systematic uncertainties in the
extracted values that result from the crude modeling of the early and
late non-equilibrium stages.

This insight is not new, and it has spurred the development of hybrid
algorithms that describe different stages of the expansion with different
tools for the last ten years. The main advantage of a hydrodynamic description
(where valid) over kinetic theory is its relative simplicity: only a few
macroscopic fields (energy and baryon density, pressure and flow velocity)
must be evolved in space-time, whereas the microscopic approach requires
to follow the evolution of both the momenta and positions of all particles.
The phenomenological success of hydrodynamic modeling for heavy-ion
collisions thus leads naturally to the concept of using a fluid dynamical
description as the backbone of the complete evolution model, interfaced with
computationally more demanding microscopic algorithms to describe the early
and late non-equilibrium stages. For the hadronic rescattering phase
several microcopic algorithms that solve coupled Boltzmann equations for
the hadron distribution functions were developed and extensively tested in
the 1980s and '90s \cite{Aichelin:1986wa,Maruyama:1993jb,Fang:1994cm,%
Bass:1998ca,Nara:1999dz}. Hybrid codes that coupled an ideal fluid
dynamical description of an expanding QGP to such hadronic rescattering
codes and compared the results with purely fluid dynamical calculations
began to appear about ten years ago \cite{Bass:2000ib,Teaney:2000cw,%
Hirano:2005xf,Nonaka:2006yn}. Here we present the first hybrid model that
interfaces {\em viscous} relativistic hydrodynamics (specifically the
(2+1)-dimensional algorithm \VISH\ \cite{Song:2007fn}) with a
hadronic Boltzmann cascade (specifically the Monte Carlo algorithm
\U\ \cite{Bass:1998ca}), via the ``hydro-to-OSCAR'' converter
\con\ that converts hydrodynamic output into hadrons, with positions
and momenta given in {\tt OSCAR}
format\footnote{http://karman.physics.purdue.edu/OSCAR-old/}
that can be read by {\tt UrQMD}. We call this hybrid code
\VC.\footnote{``{\bf V}iscous {\bf I}srael-{\bf S}tewart {\bf H}ydrodynamics
                a{\bf N}d {\bf U}rQMD''.}
In the present version of \VC\ the pre-equilibrium stage is not yet
described dynamically, but continues to be replaced by initial conditions
for the fluid dynamic evolution.

An essential ingredient in \VC\ is the use of a state-of-the-art equation of
state (EOS) for hot QCD matter, s95p-PCE \cite{Huovinen:2009yb,Shen:2010uy},
which incorporates our best knowledge of the relation between pressure,
energy and entropy density in the deconfined QGP stage from Lattice QCD
\cite{Bazavov:2009zn} (see also \cite{Borsanyi:2010zh}) and matches it to
a realistic hadron resonance gas (HRG) at low temperatures, taking into
account that the abundances of stable hadrons (after strong decays of
unstable resonances) are experimentally known to freeze out at
$\Tchem{\,\approx\,}165$\,MeV \cite{BraunMunzinger:2001ip}. This requires
the introduction of properly adjusted, temperature dependent non-equilibrium
chemical potentials for all hadronic species below $\Tchem$
\cite{Teaney:2002aj,Hirano:2002ds,Kolb:2002ve,Huovinen:2007xh}. Chemical
freeze-out is immediate\footnote{It is a consequence of the smallness of
  particle-changing inelastic cross sections when compared with the much
  larger elastic and (resonance dominated) quasielastic scattering cross
  sections that keep the hadron gas close to thermal (but not chemical)
  equilibrium for a range of temperatures below $\Tchem$
  \cite{Bravina:1998pi}.}
and automatic in \U, due to the rapid three-dimensional expansion
of the fireball in the hadronic stage \cite{Bass:1999tu}. This implies
that the hydrodynamic output that is fed into \U\ must have the
correct chemical composition, since otherwise the final hadron abundances
disagree with experiment. If we decrease the switching temperature
$\Tsw$ for the transition from hydrodynamics to hadron cascade below
the chemical freeze-out temperature $\Tchem$, the hydrodynamic output
thus has to reflect the correct partial chemical equilibrium (PCE)
abundances that would, after resonance decays, produce the correct
final yield ratios. This is also important for the elliptic flow:
Although the non-equilibrium chemical potentials have a very small
effect on the equation of state $p(e)$ \cite{Hirano:2002ds}, the
distribution of the total momentum anisotropy among hadron species
strongly depends on the chemical composition of the hadronic system
\cite{Hirano:2002ds,Kolb:2002ve,Huovinen:2007xh,Hirano:2005wx}. A
comparison between chemical equilibrium and partial chemical equilibrium
EOS in ideal hydrodynamics shows that the breaking of chemical
equilibrium in the HRG can increase the differential elliptic flow
$v_2(p_T)$ for pions by 25\% \cite{Hirano:2002ds,Huovinen:2007xh}.

\VC\ was developed to remove uncertainties from non-equilibrium
dynamics during the late hadronic stage when extracting QGP
transport coefficients from experimentally measured final hadron
spectra. The main goals of the present article, however, are purely
conceptual: by comparing the hadron spectra and differential elliptic
flow from \VC\ with those from purely hydrodynamic simulations with
different (temperature dependent) values for the shear viscosity to
entropy density ratio $\eta/s$ during the HRG stage, we want to establish
to what extent the microscopic \U\ dynamics of the HRG phase can be
mimicked by an effective macroscopic calculation based on viscous
fluid dynamics. By varying the switching temperature between \VISH\ and
\U\ we explore the existence of a ``switching window'', i.e. of a range
of temperatures within which both \VISH\ (with an appropriate choice of
transport coefficients) and \U\ provide a valid description of the fireball
evolution. Our goal here is {\em not} to compare \VC\ with experimental data
from RHIC; this is done elsewhere \cite{Song:2010mg}. Consequently, we
explore the sensitivity of our results to variations in the hydrodynamic
initial conditions only to the extent that they affect the answer to
the above questions.

The paper is organized as follows. In Sec.~\ref{sec2} we briefly present
the three components of \VC: the viscous hydrodynamic algorithm \VISH,
the hydro-to-micro converter \con, and the hadron cascade \U. In
Sec.~\ref{sec3} we compare basic observables (spectra and elliptic flow)
from \VC\ and pure viscous hydrodynamics, using both chemical equilibrium
and PCE equations of state as input. In Sec.~\ref{sec4} we test the
switching temperature dependence of the final spectra and elliptic flow
obtained with \VC, extracting an effective viscosity $(\eta/s)(T)$
for \U\ under dynamical conditions provided by RHIC collisions. We interpret
our findings and summarize our results in Sec.~\ref{sec5}. Appendix~\ref{appa}
describes tests of the hydro-to-micro converter \con\ and shows some
results that demonstrate its accuracy.

%%%%%%%%%%%%%%%%%%%%%%%%%%%%%%%%%%%%%%%%%%%%%%%%%%%%%%%%%%%%%%%%%%%%%%%%%%%
\section{\VC: coupling viscous hydro\-dynamics to a hadron cascade}
\label{sec2}
%%%%%%%%%%%%%%%%%%%%%%%%%%%%%%%%%%%%%%%%%%%%%%%%%%%%%%%%%%%%%%%%%%%%%%%%%%%

In this section we describe the structure of \VC, a hybrid code that
couples the viscous hydrodynamic expansion of the QGP stage in heavy-ion
collisions to a microscopic kinetic evolution of the dilute late hadronic
stage using a Boltzmann Monte Carlo approach. The hydrodynamic component
allows for ideal fluid evolution in the limit of zero transport coefficients.
We here include only shear viscosity, neglecting bulk viscous contributions
from the QGP and hadronization stages whose effects on spectra and elliptic
flow are expected to be small \cite{Song:2009je,Denicol:2010tr}.\footnote{Bulk
  viscous effects from the hadron resonance gas stage \cite{Bozek:2009dw} are
  automatically accounted for by the Boltzmann cascade component of the
  hybrid code.}
We assume zero net baryon density everywhere and thus do not follow
explicitly the evolution of the baryon current. The latter can be easily
included in future versions of the code.

The viscous hydrodynamic and microscopic kinetic algorithms are interfaced
with each other through the Monte Carlo event generator \con\ that converts
hydrodynamic output on a hypersurface of constant temperature $\Tsw$ to
particles, by sampling the corresponding Cooper-Frye \cite{Cooper:1974mv}
phase-space distributions, including viscous corrections. This procedure
requires switching temperatures $\Tsw$ that are low enough for \U\ to
provide a reliable description of the subsequent hadronic rescattering
dynamics. The highest possible switching temperature is therefore
$\Tsw\eq\Tc$ where $\Tc$ denotes the (pseudo)critical temperature for
the quark-hadron phase transition. Higher $\Tsw$ values would require a
microscopic description of the hadronization process itself, including
accompanying changes in vacuum structure, and knowledge of the effective
particle degrees of freedom during this process. This is at present an
unsolved problem.

For comparison we also perform purely hydrodynamic simulations without
the hadronic cascade, by running \VISH\ to lower temperatures
and decoupling directly into free-streaming particles, by using the
Cooper-Frye prescription at $\Tdec$. These comparison runs are done
with constant or temperature-dependent specific shear viscosities $\eta/s$
in the hadronic phase. For the QGP phase we use constant $\eta/s$,
motivated the weak (logarithmic) temperature dependence of $\eta/s$
predicted by perturbative QCD in the weak-coupling limit \cite{Arnold:2003zc}
and the temperature independence of $\eta/s$ predicted by the AdS/CFT
correspondence in the strong coupling limit
\cite{Policastro:2001yc}.\footnote{We acknowledge that near $\Tc$ QCD is
  not a conformal field theory and that therefore in the temperature
  region accessible at RHIC ($\Tc{\,\leq\,}T{\,\alt\,}2\Tc$) $\eta/s$
  might very well feature a stronger temperature dependence than predicted
  by both perturbative QCD and the strong coupling limit for conformal field
  theories. Heavy-ion experiments at the LHC are expected to shed light on
  this question.}

In the following subsections we discuss the components of \VC\ in
more detail.

%%%%%%%%%%%%%%%%%%%%%%%%%%%%%%%%%%%%%%%%%%%%%%%%%%%%%%%%%%%%%%%%%%%%%%%%%%%%%
\subsection{Viscous hydrodynamics (\VISH)}
\label{sec2a}
%%%%%%%%%%%%%%%%%%%%%%%%%%%%%%%%%%%%%%%%%%%%%%%%%%%%%%%%%%%%%%%%%%%%%%%%%%%%%

For the hydrodynamic stage we use \VISH, a (2{+}1)-dimensional viscous
hydrodynamic code with longitudinal boost invariance \cite{Song:2007fn}.
The specific implementation used in the present work, including the
equation of state s95p-PCE, is described in Sections II and III of
Ref.~\cite{Shen:2010uy} to which we refer the reader for technical details.
For later reference we note that the relaxation time for the shear pressure
tensor is set to $\tau_\pi\eq3\eta/(sT)$. In some of the comparison runs
we also employ the equation of state SM-EOS~Q described in
\cite{Song:2007fn} which assumes a first-order quark-hadron
phase transition and chemical equilibrium (CE) in the hadronic phase.
The comparison between pure hydrodynamic simulations with s95p-PCE and
SM-EOS~Q (CE) emphasizes the effects arising from the breaking of chemical
equilibrium in the hadronic phase at temperatures below $\Tchem\eq165$\,MeV.

Different from Ref.~\cite{Shen:2010uy}, we here use as default initial
conditions an initial energy density profile taken to be 100\% proportional
to the wounded nucleon density from the optical Glauber model whose peak
energy density in central ($b\eq0$) Au+Au collisions is normalized to
$e_0{\,\equiv\,}e(r{=}0,\tau_0;b{=}0)\eq30$\,GeV/fm$^3$, with
$\tau_0\eq0.6$\,fm/$c$. This gives roughly the correct final charged
hadron multiplicity $\dNdy$ in central collisions, but does not correctly
reproduce its measured nonlinear dependence on the total number of wounded
nucleons $\Npart$ in noncentral collisions. For the purposes of the present
conceptual study this is not essential.

In Sec.~\ref{sec4}, in order to test the ``universality'' of the effective
temperature-dependent $(\eta/s)(T)$ for \U\ extracted by comparing
\VC\ with \VISH, we also use Color Glass Condensate (CGC) motivated
initial conditions, obtained by averaging a large number of fluctuating
initial entropy density profiles computed with the {\tt fKLN} Monte Carlo
code\footnote{Available at URL \newline
  [http://th.physik.uni-frankfurt.de/\~{}drescher/CGC/]}
\cite{Drescher:2006ca} after recentering them and aligning their major
axes.\footnote{The CGC initial profiles were provided by T. Hirano
  \cite{Hirano:2009ah}. The same profiles were used in \cite{Song:2010mg}.}
The resulting smooth average entropy density is converted to an initial energy
density profile using the EOS. This procedure accounts, in an average way,
for event-by-event fluctuations in the initial source eccentricity, giving
(depending on impact parameter) $\sim 30-100\%$ larger initial
eccentricities than the optical Glauber model (see \cite{Hirano:2009ah}
for details). For both optical Glauber and fluctuating {\tt fKLN} initial
profiles we assume zero transverse flow at the beginning of the hydrodynamic
evolution at $\tau_0$.

%%%%%%%%%%%%%%%%%%%%%%%%%%%%%%%%%%%%%%%%%%%%%%%%%%%%%%%%%%%%%%%%%%%%%%
\subsection{Hydro-to-micro converter (\con)}
\label{sec2b}
%%%%%%%%%%%%%%%%%%%%%%%%%%%%%%%%%%%%%%%%%%%%%%%%%%%%%%%%%%%%%%%%%%%%%%

To convert the hydrodynamic output into particles that can then be further
propagated with the hadron cascade \U\ we first use the {\tt AZHYDRO}
algorithm to find an isothermal freeze-out surface $\Sigma(x)$ of
temperature $\Tsw$ and then calculate the hadron spectra with the
Cooper-Frye formula~\cite{Cooper:1974mv} (see \cite{Song:2007fn} for
details). \con\ is a Monte Carlo event generator that samples tiles on
$\Sigma$ and generates the position $x$ and momentum $p$ of a particle
of species $i$ with a probability derived from the differential Cooper-Frye
formula:
\begin{eqnarray}
\label{Cooper}
  E\frac{d^3N_i}{d^3p}(x) &=& \frac{g_i}{(2\pi)^3}
  p\cdot d^3\sigma(x)\, f_i(x,p)
\\
  &=& \frac{g_i}{(2\pi)^3} p\cdot d^3\sigma(x)
 \left[f_{\mathrm{eq},i}(x,p)+\delta f_i(x,p) \right].
\nonumber
\end{eqnarray}
Here $g_i$ is the degeneracy factor for hadron species $i$, and
$d^3\sigma_\mu(x)$ is the outward-pointing surface normal vector
of the selected tile at point $x$ on the surface $\Sigma$.
$f_i(x,p)\eq{f}_{\mathrm{eq},i}(x,p)+\delta f_i(x,p)$ is the local
distribution function for hadron species $i$, consisting of a local
equilibrium part (here with $\mu=0$)
\begin{eqnarray}
\label{f0}
  f_{\mathrm{eq},i}(x,p)
  = \frac{1}{e^{p\cdot u(x)/T(x)}\pm 1}
\end{eqnarray}
and a (small) deviation $\delta f_i$ from local equilibrium due to shear
viscous effects. For $\delta f$ we make the quadratic ansatz (see
\cite{Teaney:2003kp,Baier:2006um} for a discussion of this and other
possibilities)
\begin{eqnarray}
\label{deltaf}
  \!\!\!\!
  \delta f(x,p)\!\!&=&\!\!
  f_\mathrm{eq}(p,x) \bigl(1{\mp}f_\mathrm{eq}(p,x)\bigr)
  \frac{p^\mu p^\nu \pi_{\mu\nu}(x)}{2T^2(x)\left(e(x){+}p(x)\right)}\
%\nonumber\\
\end{eqnarray}
(the upper (lower) sign is for fermions (bosons)). $\delta f$ is
proportional to the shear viscous pressure tensor $\pi^{\mu\nu}(x)$
on the freeze-out surface and increases (in our case) quadratically with
the particle momentum. In the limit $\eta/s{\,\to\,}0$ (i.e. for ideal fluid
dynamics) $\delta f$ vanishes.

Since \VISH\ is a (2+1)-dimensional code with longitudinal boost
invariance but \U\ propagates particles in all three spatial dimensions,
we extend the \VISH\ output analytically from midrapidity ($y\eq0$)
to non-zero momentum rapidities, using boost invariance. After having
determined the transverse position and transverse momentum of a particle
using a space-time rapidity integrated version of Eq.~(\ref{Cooper}) as
weight, we sample its momentum rapidity $y$ randomly within the finite
range $-3{\,<\,}y{\,<\,}3$, with a sharp cutoff at its upper and lower
ends. (We restricted the range to $|y|{\,<\,}3$ to minimize the excitation
of strings in \U.) Its space-time rapidity $\eta_s$ (defining its
longitudinal position)
is then sampled according to the $\eta_s$-dependence of Eq.~(\ref{Cooper})
(see Eq.~(\ref{A1}) for an explicit expression). This results in
an $\eta_s$ distribution of the generated particles that is flat
near $\eta_s\eq0$ but smeared at the edges around $\eta_s\eq\pm3$ (see
inset in Fig.~\ref{F15} for illustration). \U\ propagates the resulting
particles in all three spatial directions but (due to the nearly
boost-invariant input over the $y$- and $\eta_s$-range mentioned)
preserves boost invariance of the final momentum distributions accurately
within the region $|y|{\,<\,}1.5$. This allows comparison of the calculations
with midrapidity data from the RHIC experiments without having to worry
about edge effects from the rapidity cutoff in \con.

The default switching temperatures used in this article are $\Tsw\eq165$\,MeV
for s95p-PCE and  $\Tsw\eq160$\,MeV for SM-EOS~Q\,(CE). Here, $165$\,MeV
is equal to the chemical freeze-out temperature $\Tchem$ used in s95p-PCE,
and $160$\,MeV is the transition temperature from the mixed to hadronic
phase in SM-EOS~Q \cite{Song:2007fn}. In Sec.~\ref{sec4} we also use lower
$\Tsw$ values in order to study the existence of a switching window for \VC.

\begin{figure*}[t]
\centering
\vspace{-15mm}
\includegraphics[width=0.75\linewidth,clip=,angle=270]{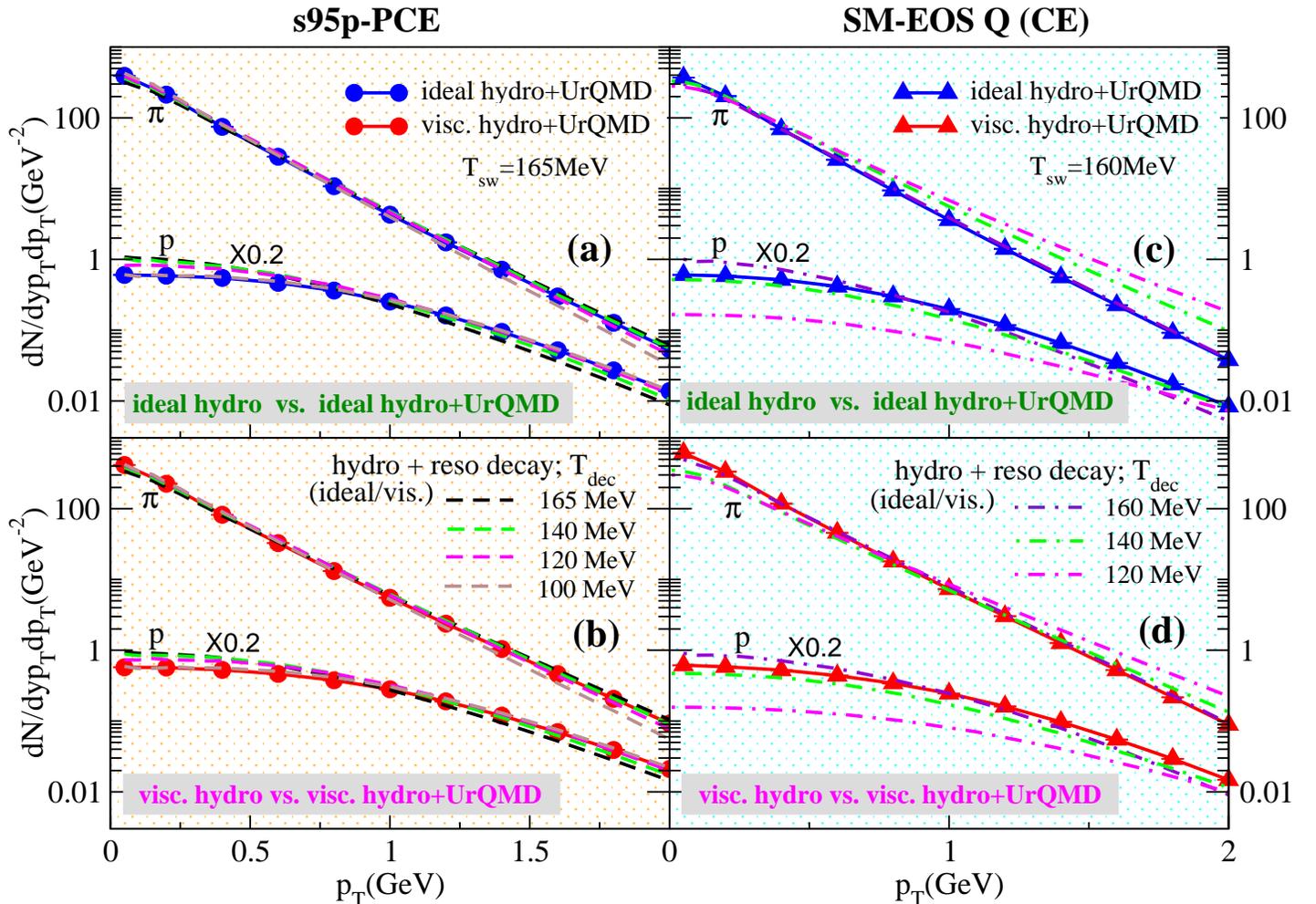}
\vspace{-7mm}
\caption{\label{F1} (Color online) Pion and proton $p_T$-spectra for
$b\eq7$\,fm Au+Au, calculated from \VISH\ and \VC with s95p-PCE (left
panels) and SM-EOS~Q(CE) (right panels). Upper panels compare pure ideal
fluid dynamics and \VC\ with ideal fluid input; lower panels compare pure
viscous hydrodynamics and \VC\ with viscous hydrodynamic input, using
$\eta/s=0.08$ in both cases during the fluid dynamic stage. Identical
Glauber model initial conditions are used in all runs
($e_0\eq30$\,GeV/fm$^3$ at $\tau_0\eq0.6$\,fm/$c$).}
\end{figure*}

For each hydrodynamic run we use \con\ to generate a large number of
events\footnote{Sufficient accuracy for the $p_T$-integrated $v_2$ is
   obtained with event samples ranging from 2,000 for 0-5\% centrality
   to 72,000 for 70-80\% centrality. For Figs.~\ref{F1}--\ref{F3} we
   used 90,000 events in order to obtain sufficient statistics for
   $v_2(p_T)$ of protons out to $p_T\eq2$\,GeV/$c$.}
whose particles are then further propagated with \U\ until all collisions
cease and unstable resonances have decayed. In the purely hydrodynamic
simulations we run the converter on the final decoupling surface with
temperature $\Tdec$ to generate events of free-streaming hadrons with
similar statistics. (We use $\Tdec\eq100$\,MeV with EOS s95p-PCE and
$\Tdec\eq130$\,MeV with SM-EOS~Q\,(CE), corresponding to similar
decoupling energy densities \cite{Hirano:2002ds}.) In these runs we have
no further hadronic collisions and only allow the unstable resonances to
decay, and analyze the final state with the same tools as used for
complete \VC\ runs.

%%%%%%%%%%%%%%%%%%%%%%%%%%%%%%%%%%%%%%%%%%%%%%%%%%%%%%%%%%%%%%%%%%%%%%%%%%%%
\subsection{Microscopic hadronic transport (\U)}
\label{sec2c}
%%%%%%%%%%%%%%%%%%%%%%%%%%%%%%%%%%%%%%%%%%%%%%%%%%%%%%%%%%%%%%%%%%%%%%%%%%%%

For the modeling of the hadronic phase, we use the Ultra-relativistic
Quantum Molecular Dynamics (\U) model \cite{Bass:1998ca}, a microscopic
hadronic transport model based on the Boltzmann equation. \U, which was
initially developed as an ab-initio model for the simulation of
relativistic heavy-ion collisions, is well suited for the description
of a hadron gas in and out of chemical equilibrium \cite{Bravina:1998pi,%
Belkacem:1998gy,Bravina:2000dk,Demir:2008tr} and has been successfully
applied to previous hybrid micro+macro transport approaches based on
ideal fluid dynamics \cite{Bass:2000ib,Nonaka:2006yn,Werner:2009zza}.

In \U, the system evolves through a sequence of binary collisions and
$2{\,\to\,}N$ decays of mesons and baryons. The cross sections are assumed
to be free vacuum cross sections and depend on the center of mass energy
of the two colliding hadrons as well as on their flavor and quantum numbers.
The \U\ collision term contains 49 different baryon species (including
nucleon, delta and hyperon resonances with masses up to 2 GeV) and 25
different meson species (including strange meson resonances), which
are supplemented by their corresponding anti-particle and all
isospin-projected states. Full baryon/antibaryon symmetry is included.
For excitations with higher masses a string picture is used.
All states listed can be produced in string decays, s-channel
collisions or resonance decays.

Tabulated or parameterized experimental cross sections are used when
available, resonance absorption and scattering is handled via the
principle of detailed balance. If no experimental information is
available, the cross section is either  calculated via an OBE model
or via a modified additive quark model which takes basic phase space
properties into account. A detailed overview of the elementary cross
sections and processes included in the \U\ model is
given elsewhere \cite{Bass:1998ca}.

When modeling the hadronic phase subsequent to the decay of a thermalized
QGP, we find that the relative momenta and c.m. energies of the individual
hadron-hadron interactions are rather small and therefore
string excitations and decays are strongly suppressed and occur rarely.
The dominant forms of interactions encountered are elastic scattering
and inelastic scattering through resonance formation and decay.

It is important to note that \U\ makes no equilibrium assumptions and
can therefore be utilized for the description of systems in and out of
equilibrium. \U\ will retain equilibrium if given an equilibrium initial
condition with suitable boundary conditions. This characteristic allows
for the construction of our hybrid approach, since it ensures that \U's
response to the microscopic hadronic configuration generated by \con\ will
initially mimic the response of the viscous hydrodynamic model. However,
the real advantage of transitioning to a microscopic transport such as
\U\ lies in its ability to describe the evolution of systems out of
equilibrium, e.g. prior to equilibration or during the break-up and
freeze-out stage of the reaction when assumptions of chemical and/or
kinetic equilibrium are no longer valid.

%============================ Fig. 1 ==================================
%======================================================================

By virtue of its microscopic nature, \U\ takes the full local temperature
and particle fugacity dependence of the hadronic viscosity into account
\cite{Demir:2008tr}, even though it may be challenging to quantify the
exact value of the hadronic viscosity inherent in the \U\ calculation of
the hadronic phase (due to its dependence not only on temperature but
also on multiple (non-equilibrium!) particle fugacities).

%%%%%%%%%%%%%%%%%%%%%%%%%%%%%%%%%%%%%%%%%%%%%%%%%%%%%%%%%%%%%%%%%%%%%%%%%%%%%
\section{Spectra and flow from hydrodynamics and the hybrid model}
\label{sec3}
%%%%%%%%%%%%%%%%%%%%%%%%%%%%%%%%%%%%%%%%%%%%%%%%%%%%%%%%%%%%%%%%%%%%%%%%%%%%%
\subsection{EOS dependence: s95p-PCE vs. SM-EOS~Q}
\label{sec3a}
%%%%%%%%%%%%%%%%%%%%%%%%%%%%%%%%%%%%%%%%%%%%%%%%%%%%%%%%%%%%%%%%%%%%%%%%%%%%%

In this subsection we calculate transverse momentum spectra and elliptic
flow for identified hadrons in the hybrid approach and compare them to
pure ideal and viscous hydrodynamic calculations, focusing in particular
on the EOS and the $\Tdec$ dependences of the calculations. The
main idea of this comparison is to determine whether the results of the
hybrid model calculation can be reproduced by pure ideal or viscous
hydrodynamics, or if specific features in the spectra and the transverse
momentum or centrality dependence of $v_2$ lead to clear discriminators
between the two approaches. Unless otherwise noted, we use initial
conditions and switching and decoupling parameters as described in
Sec.~\ref{sec2}. With identical initial conditions and identical transport
coefficients $\eta/s$ in the hydrodynamic QGP stage, we ensure that the
only difference between the \VISH\ and \VC\ runs lies in the treatment
of the hadronic stage. In the SM-EOS~Q(CE) case this means that the
pure hydrodynamic runs assume both thermal and chemical equilibrium in
the hadronic phase, whereas the \U\ component of \VC\ breaks both. When
we use \VISH\ with s95p-PCE, we ensure that the pure hydrodynamic and
\VC\ hybrid simulations use the same non-equilibrium chemical composition
in the hadronic stage; in this case the difference lies only in the
assumption of (approximate) thermal equilibrium for the hydrodynamic
runs whereas the \U\ cascade in \VC\ allows the system to evolve far
out of local thermal equilibrium, all the way to final decoupling.

%============================ Fig. 2 ==================================
\begin{figure*}[htb]
\vspace{-0mm}
\includegraphics[width=0.35\linewidth,clip=,angle=270]{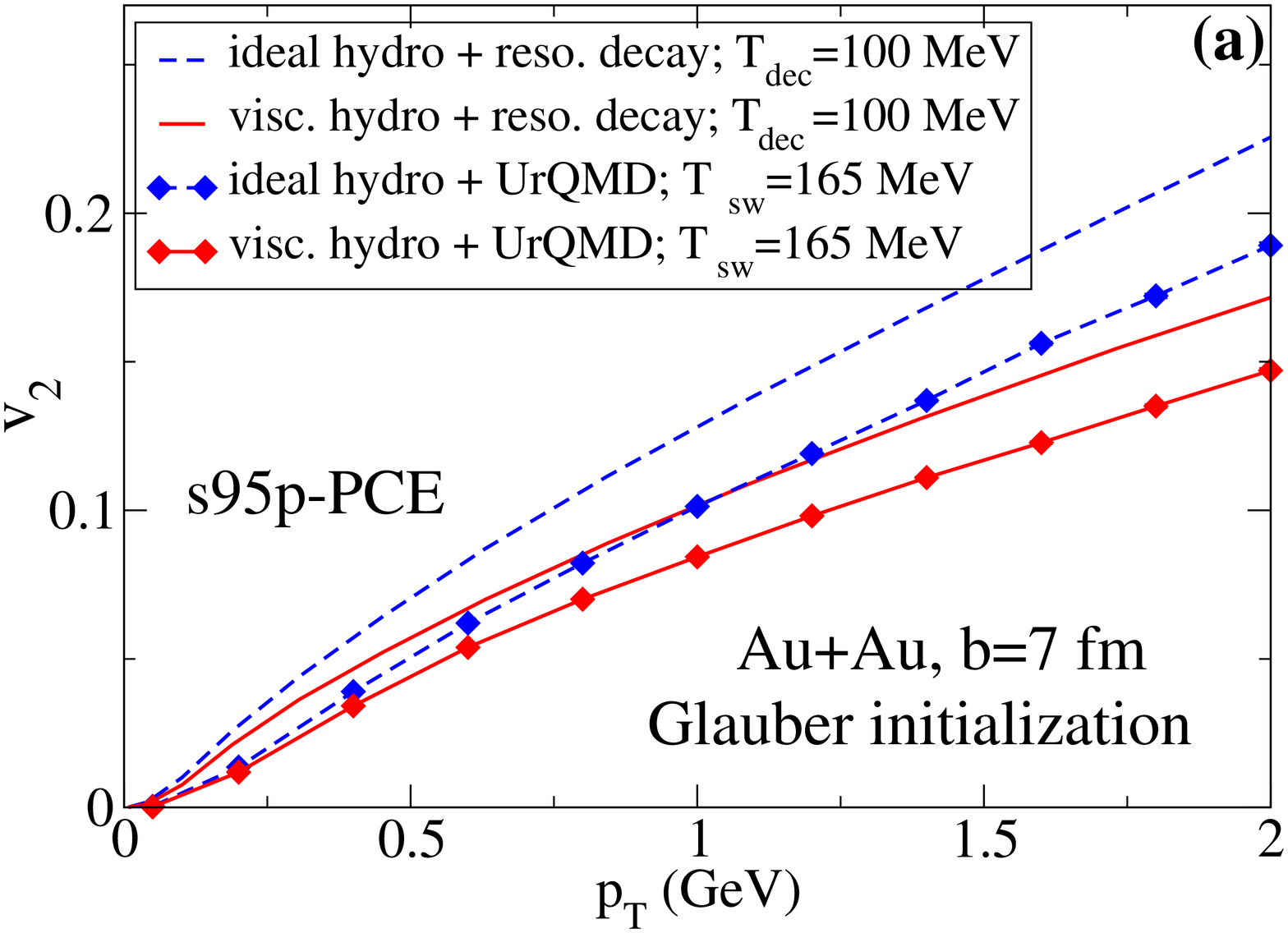}
\includegraphics[width=0.35\linewidth,clip=,angle=270]{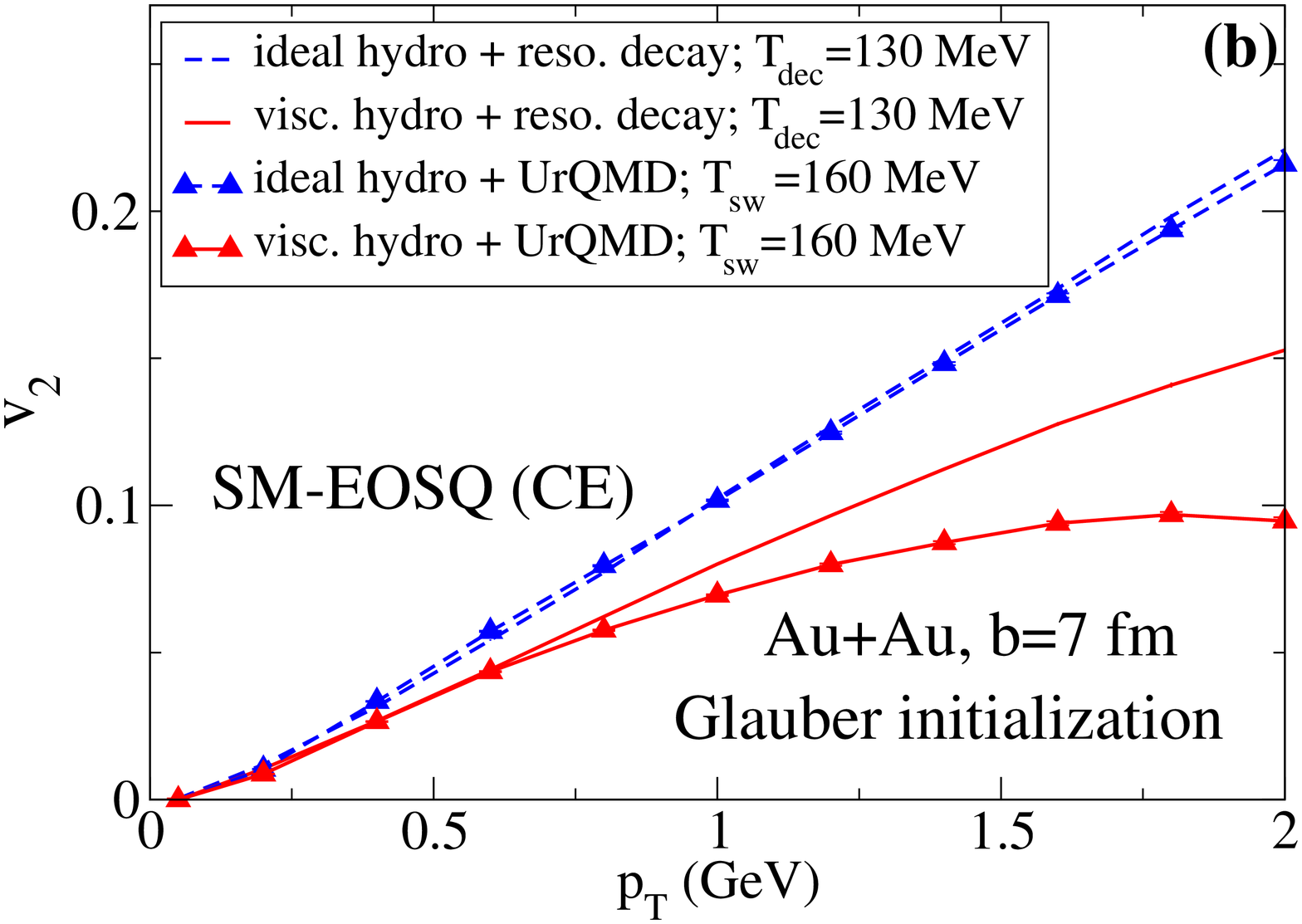}
\caption{\label{F2} (Color online) $v_2 (p_T)$ for all hadrons from
$b\eq7$\,fm Au+Au collisions, calculated from \VISH\ (lines without symbols)
and \VC\ (lines with symbols). Dashed and solid lines are for ideal
($\eta/s\eq0$) and viscous ($\eta/s\eq0.08$) fluids during the hydrodynamic
stage, with EOS s95p-PCE (a) and SM-EOSQ~Q(CE) (b), respectively.
Same initial conditions as in Fig.~\ref{F1}.}
\end{figure*}
%======================================================================

The pion and proton transverse momentum spectra shown in the left panels of
Fig.~\ref{F1} demonstrate good agreement between the pure hydrodynamic and
hybrid \VC\ runs as long as EOS s95p-PCE is used, i.e. as long as we ensure
that the hydrodynamic simulations correctly implement the non-equilibrium
chemical evolution in the hadronic phase. With the correct PCE equation of
state, the hydrodynamic pion spectra are almost insensitive to the choice
of kinetic decoupling temperature $\Tdec$, agreeing well with their
counterparts from the hybrid model \VC\ in all cases. The hydrodynamic
proton spectra become flatter as $\Tdec$ decreases, due to build-up of
additional radial flow, and the best agreement with the hybrid model is
achieved for the lowest value shown in the graph ($\Tdec\eq100$\,MeV). This
demonstrates that significant additional radial flow is generated during
the hadronic rescattering stage \cite{Teaney:2000cw,Hirano:2007ei}, but
that for pions (whose momentum distributions react less strongly to radial
flow than the heavier protons) the radial flow and cooling effects on the
spectral slope balance each other as we lower $\Tdec$ \cite{Hirano:2005wx}.
Figure~\ref{F1} shows that all of the above statements hold irrespective
of whether we assume zero or non-zero viscosity during the QGP stage: in
both cases the transverse momentum spectra from the hybrid code \VC\ can
be well represented by purely hydrodynamic simulations with
$\Tdec\eq100$\,MeV and $\eta/s$ values that do not change between the
QGP and hadron gas stages.

The right panels in Fig.~\ref{F1} show that the same is not true when we
use SM-EOS~Q in the hydrodynamic code which assumes chemical equilibrium
in the hadronic stage, contrary to the underlying microscopic \U\ dynamics.
In this case agreement between the pion spectra from \VC\ and the
purely hydrodynamic simulations requires immediate decoupling at
$\Tsw\eq160$\,MeV, but the corresponding purely hydrodynamic proton
spectra are too steep because they lack the boost from the additional
radial flow developed by \U\ during the hadronic stage. Lowering $\Tdec$
in the pure hydrodynamic runs helps with the shape of the proton spectra,
but quickly eats into the total proton yield (i.e. the normalization
of the proton spectra), due to annihilation with antiprotons, and
simultaneously the pion spectra become too flat when compared with
\VC.

We conclude that, at the level of single-particle $p_T$ spectra, it
is possible to simulate the full microscopic dynamics of the \U\ hadron
cascade by viscous (or even ideal) fluid dynamics as long as a PCE EOS
is used that correctly describes the non-equilibrium chemical composition
in \U, and one allows for the buildup of additional hadronic radial flow
by setting a low decoupling temperature $\Tdec{\,\simeq\,}100$\,MeV.
With SM-EOS~Q(CE), purely hydrodynamic simulations are unable to reproduce
the \VC\ spectra for any choice of $\Tdec$; for the convenience of the
following academic comparison of elliptic flow $v_2$ within these two
approaches, we choose for SM-EOS~Q the "historical standard value"
$\Tdec\eq130$\,MeV \cite{Kolb:1999it}.

Research over the past few years has established that elliptic flow
$v_2$ is influenced by both hadronic dissipative effects
\cite{Hirano:2005xf} and the chemical composition of the hadronic matter
\cite{Huovinen:2007xh}. The reaction dynamics modeled by \U\ contains
both types of non-equilibrium effects. To isolate the effect of kinetic
non-equilibrium we can compare results from \VC\ with pure hydrodynamic
calculations that model the same hadronic chemical non-equilibrium
composition as \U, by using EOS s95p-PCE. This is shown in Fig.~\ref{F2}(a),
for both an ideal ($\eta/s\eq0$) and minimally viscous $\eta/s\eq0.08$
QGP fluid. For $(\eta/s)_\mathrm{QGP}\eq0$ one finds that in the hybrid
model calculations $v_2$ is suppressed by $\sim 15\%$ at $p_T\eq2$\,GeV,
as a result of the hadronic viscosity inherent in the \U\ dynamics.
For $\eta/s\eq0.08$ the difference between the purely hydrodynamic and
hybrid model results is a bit less ($\sim10\%$) since now the non-zero
hydrodynamic viscosity already partially accounts for the viscous $v_2$
suppression in \U. Comparing the ideal and viscous pure hydrodynamic
results with each other we find $\sim 20\%$ viscous $v_2$ suppression
at $p_T\eq2$\,GeV for $\eta/s\eq0.08$, consistent with earlier results
\cite{Teaney:2003kp,Song:2007fn,Romatschke:2007mq}. Fig.~\ref{F2}(a)
demonstrates that evolution with the hybrid model suppresses $v_2$ more
strongly than viscous hydrodynamics alone with the same $\eta/s$;
a realistic microscopic description of dissipative hadron dynamics
within a hybrid approach is therefore essential.\footnote{We will see
below that merely adjusting the specific shear viscosity $\eta/s$ to
larger values in the hadronic phase is not sufficient.}

Figure~\ref{F2}(b) shows that an incomplete separation of chemical and
kinetic non-equilibrium effects can lead to misleading conclusions. Using
a chemical equilibrium EOS (SM-EOS~Q) in the hadronic phase and comparing
a pure ideal fluid calculation with a \VC\ simulation with ideal fluid
input (which is out of chemical equilibrium during most of the hadronic
stage) happens to yield (at least for $b\eq7$\,fm in
Au+Au\footnote{When compared with experiment, the hybrid model calculations
  exhibit a significantly improved centrality dependence compared to
  pure ideal fluid dynamics. The almost perfect cancellation of kinetic and
  chemical non-equilibrium effects on $v_2(p_T)$ seen in Fig.~\ref{F2}(a)
  may therefore not work as well at other impact parameters, although
  the tendency of the two effects to work against each other is generic
  \cite{Shen:2010uy}.}
) almost identical results for $v_2(p_T)$. Viscous suppression of $v_2$ by
kinetic non-equilibrium \cite{Hirano:2005xf} almost exactly balances
\cite{Teaney:2000cw} the previously observed enhancement of $v_2(p_T)$
caused by chemical non-equilibrium in the hadronic phase
\cite{Hirano:2002ds,Kolb:2002ve,Huovinen:2007xh}. The solid lines comparing
pure viscous hydrodyna\-mics and \VC\ with viscous fluid input demonstrate
that this cancellation is accidental and no longer occurs when describing
the QGP as a viscous fluid. In this case \VC\ gives much lower $v_2(p_T)$
than the pure viscous hydrodynamic calculation, mainly caused by large
negative contributions from the viscous correction $\delta f$ to the local
distribution function on the switching surface which the subsequent
\U\ dynamics is unable to erase. Viscous hydrodynamics with constant
$\eta/s$ in the hadronic phase, on the other hand, evolves back towards
local thermal equilibrium such that on the final decoupling surface at
$\Tdec$ only small $\delta f$ contributions remain.\footnote{This is, of
  course, an academic comparison since viscous hydrodynamics with constant
  $\eta/s$ cannot consistently account for hadronic freeze-out.}

%%%%%%%%%%%%%%%%%%%%%%%%%%%%%%%%%%%%%%%%%%%%%%%%%%%%%%%%%%%%%%%%%%%%%%%%%%%%%%
\subsection{Hadronic contribution to elliptic flow}
\label{sec3b}
%%%%%%%%%%%%%%%%%%%%%%%%%%%%%%%%%%%%%%%%%%%%%%%%%%%%%%%%%%%%%%%%%%%%%%%%%%%%%%
\vspace*{-3mm}

%============================ Fig. 3 ==================================
\begin{figure*}[t]
\includegraphics[width=0.35\linewidth,clip=,angle=270]{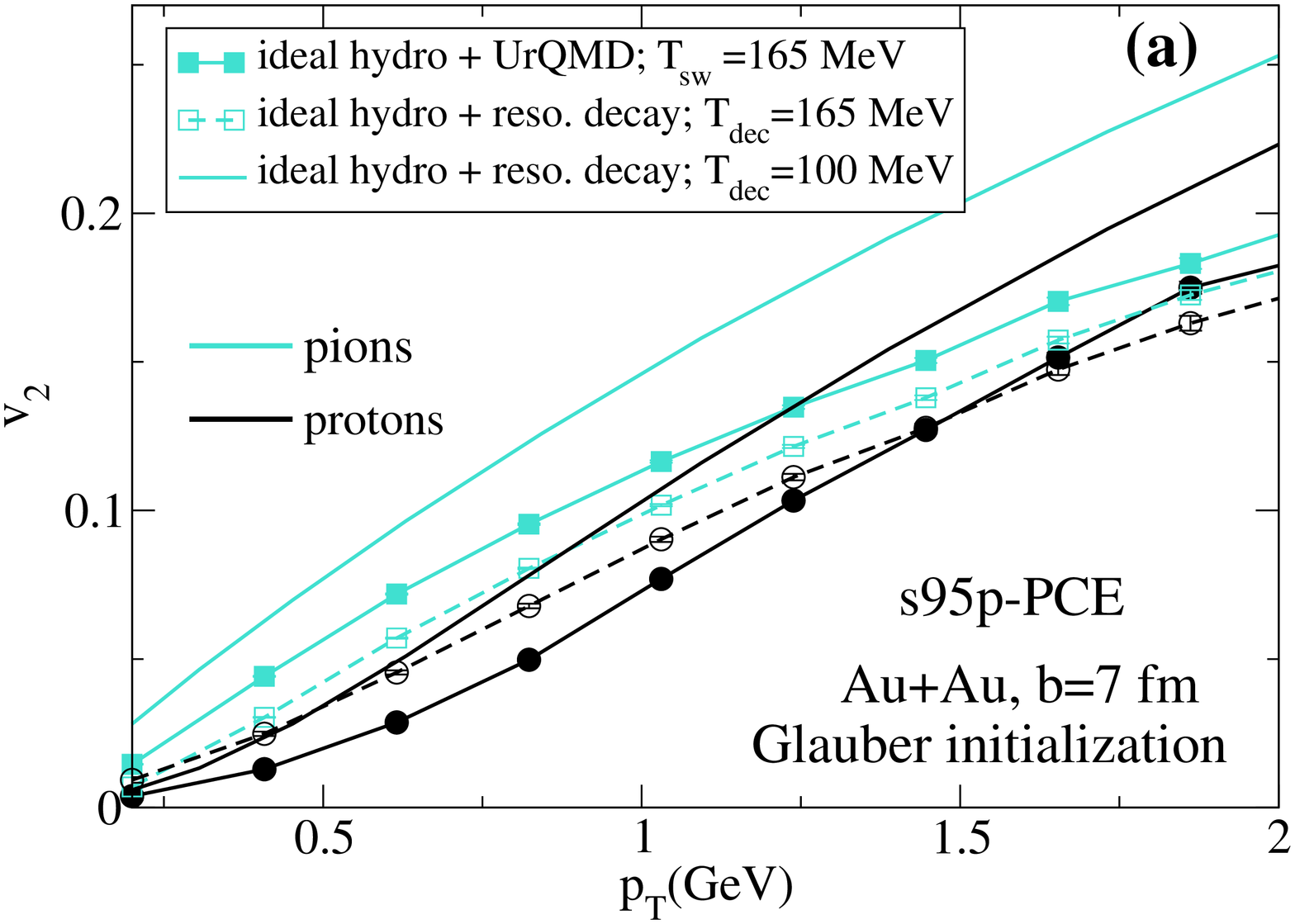}
\includegraphics[width=0.35\linewidth,clip=,angle=270]{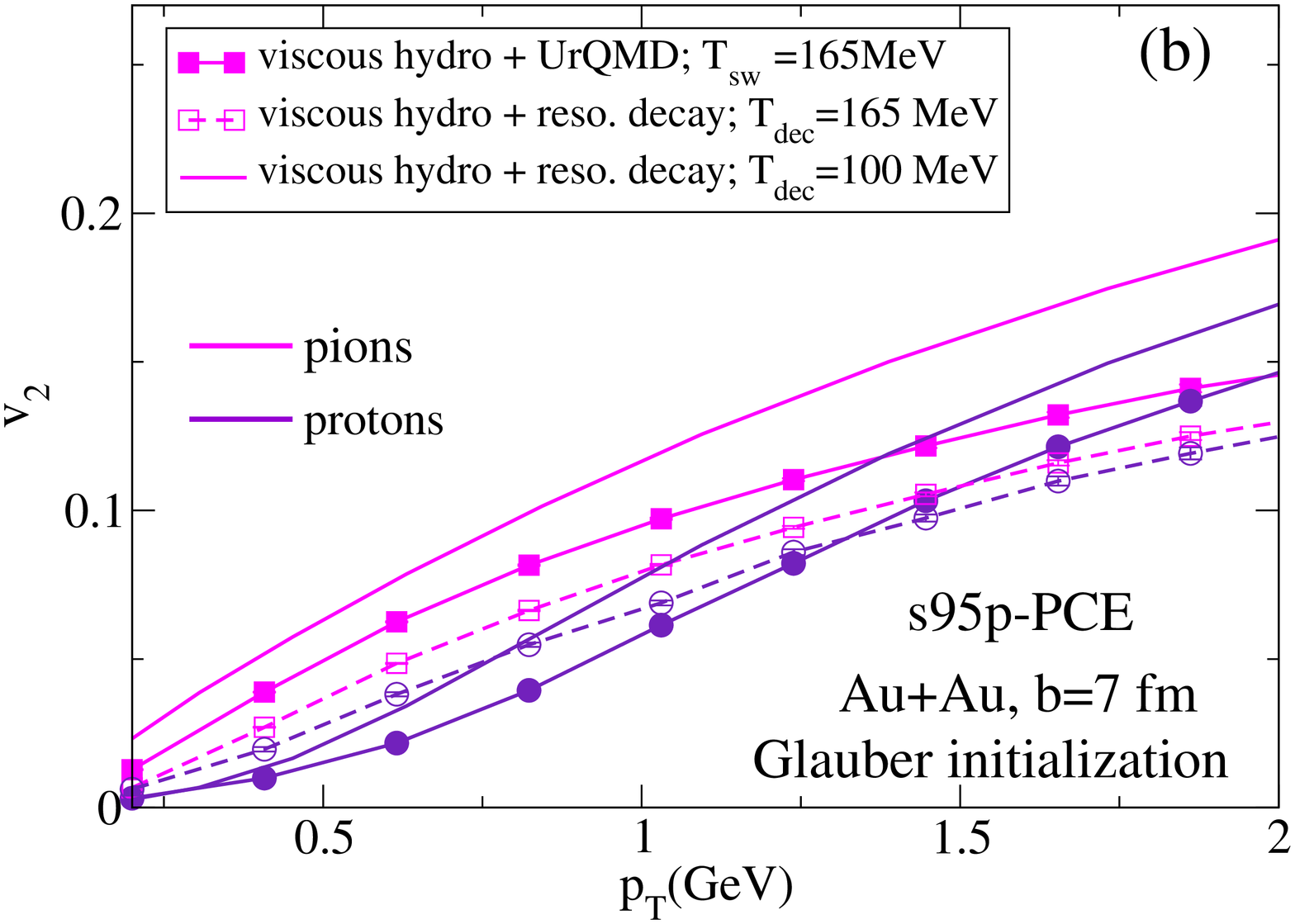}
\caption{\label{F3} (Color online) Differential elliptic flow $v_2(p_T)$
for pions (light blue/magenta) and protons (black/purple) from $b\eq7$\,fm
Au+Au collisions. Results from the hybrid model \VC (solid lines with
solid symbols) are compared with purely hydrodynamic simulations for
$\eta/s\eq0$ (a) and $\eta/s\eq0.08$ (b) for decoupling temperatures
$\Tdec\eq165$\,MeV (dashed with open symbols) and $\Tdec\eq100$\,MeV
(solid without symbols). All calculations use EOS s95p-PCE in the
hydrodynamic stage. Initial conditions are the same as in Fig.~\ref{F1}.
Please note the suppressed zeroes on the horizontal axes.}
\end{figure*}
%======================================================================

Having established that identical chemical compositions in the pure
hydrodynamic and hybrid model evolutions are essential for a meaningful
comparison that aims to assess dissipative effects in the hadronic stage,
we will from now on use the chemically frozen EOS s95p-PCE in all
simulations.

In Fig.~\ref{F3} we compare the differential pion and proton elliptic flow,
$v_2(p_T)$, from \VC\ (solid lines with symbols) with pure fluid dynamical
calculations, implementing kinetic freeze-out either directly at
$\Tsw\eq\Tchem\eq165$\,MeV (dashed lines with symbols) or at
$\Tdec\eq100$\,MeV (solid lines without symbols). The curves in panel (a)
and (b) assume $\eta/s\eq0$ and  $\eta/s\eq0.08$, respectively, during
the fluid dynamic stage. We focus our attention on the mass splitting
between pions and protons. The smallest mass splitting is found for
immediate decoupling at $\Tsw$ (dashed lines). Hadronic rescattering in
\VC\ (solid lines with symbols) increases the mass splitting at low $p_T$,
by pushing $v_2(p_T)$ up for pions and towards larger $p_T$ for protons
\cite{Hirano:2007ei}. Both of these effects strengthen if we replace
the microscopic hadronic rescattering cascade by macroscopic hydrodynamic
evolution \cite{Hirano:2007ei}. The depletion of proton $v_2$ at small
$p_T$ is a consequence of additional radial flow buildup in the hadronic
stage which pushes the heavy protons more efficiently than the light pions
to larger tranverse momenta (see Fig.~\ref{F1}); the larger $v_2(p_T)$ for
low-$p_T$ pions reflects (at least partially) a simultaneous increase of
the total momentum anisotropy, in response to the remaining spatial
fireball eccentricity that survives into the hadronic stage. These
effects are qualitatively similar for ideal (Fig.~\ref{F3}a) and viscous
fluids (Fig.~\ref{F3}b), although the larger $\delta f$ corrections
at $\Tsw$ in viscous hydrodynamics lead to an additional downward shift
of $v_2(p_T)$ for both pions and protons at large $p_T$ when hadronic
rescattering is turned off (dashed lines).

%%%%%%%%%%%%%%%%%%%%%%%%%%%%%%%%%%%%%%%%%%%%%%%%%%%%%%%%%%%%%%%%%%%%%%%%%%%%%%
\subsection{Viscous $\bm{v_2}$ suppression}
\label{sec3c}
%%%%%%%%%%%%%%%%%%%%%%%%%%%%%%%%%%%%%%%%%%%%%%%%%%%%%%%%%%%%%%%%%%%%%%%%%%%%%%

\begin{figure}[b]
\includegraphics[width=0.7\linewidth,clip=,angle=270]{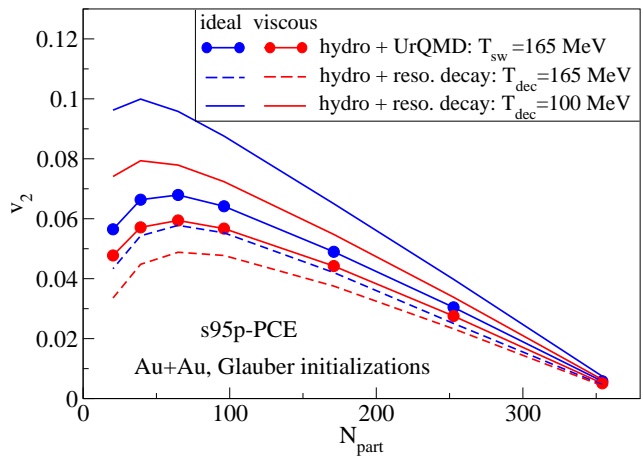}
\caption{\label{F4} (Color online) Centrality dependence of the
$p_T$-integrated elliptic flow of all hadrons, for the same parameters as
in Fig. \ref{F3}.
}
\end{figure}

Figure~\ref{F4} shows the net effect of the hadronic medium modifications
of the transverse momentum spectra and differential elliptic flow on the
$p_T$-integrated charged hadron $v_2$, as a function of collision centrality
(parametrized by the number of participants $\Npart$). The smallest
amount of $v_2$ is obtained without any hadronic rescattering at all
(dashed lines), with an additional suppression of about 20\% for the
%
%============================ Fig. 4 ==================================
%======================================================================
%
minimally viscous fluid ($\eta/s\eq0.08$) relative to the ideal fluid.
The \U\ module in the hybrid code \VC\ creates about 15\% additional
$v_2$ via hadronic rescattering, but a hydrodynamic description of the
hadron gas, treating it either as an ideal or a minimally viscous fluid,
generates a much larger hadronic contribution to $v_2$. This reflects
the fact that in the hadronic stage the fireball is still out-of-plane
elongated, and demonstrates that even a viscous fluid with $\eta/s\eq0.08$,
but especially an ideal fluid is much more efficient than \U\ in converting
this residual fireball eccentricity into additional elliptic flow.

%
%============================ Fig. 5 ==================================
\begin{figure*}[bht]
\centering
\includegraphics[width=0.33\linewidth,clip=,angle=270]{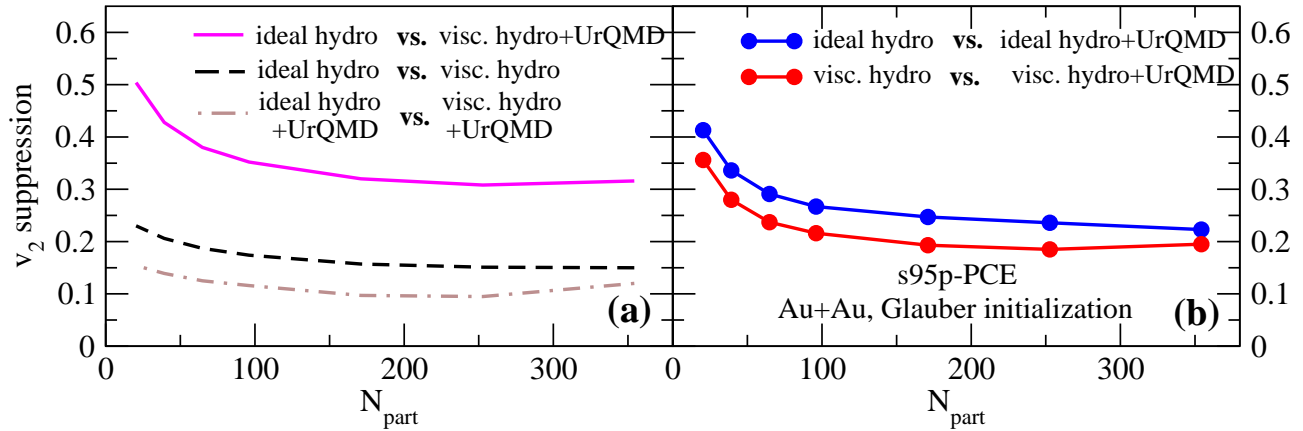}
\caption{\label{F5} (Color online) Viscous $v_2$ suppression
between different lines shown in Fig. \ref{F4}.
}
\end{figure*}
%======================================================================
%

To quantify the suppression of integrated elliptic flow by viscous
QGP and hadronic dissipation effects we define the ratio
\begin{eqnarray*}
  v_2^\mathrm{supp.} = \frac{v_2^{\mathrm{A}}-v_2^{\mathrm{B}}}
                            {v_2^{\mathrm{A}}}
\end{eqnarray*}
where A and B denote two different dynamical evolution models. Figure~\ref{F5}a
shows that viscous suppression effects are generically larger in peripheral
than in central collisions, due to the smaller fireball sizes. The solid line
in panel (a) confirms the naive expectation that the strongest suppression
(here 30--50\%, depending on collision centrality) should be seen when
comparing \VC\ with viscous fluid input to a purely hydrodynamic, ideal
fluid simulation. Among the
combinations studied in Fig.~\ref{F4}, this case maximizes the suppression
of elliptic flow by combining (in the \VC\ simulation) viscous effects
in the QGP with large dissipative effects in the hadronic stage. Replacing
the \U\ part of the evolution by viscous hydrodynamics with $\eta/s\eq0.08$
(dashed line) yields only about half of the viscous $v_2$ suppression
observed in \VC. The smallest suppression ratio ($\sim10-15\%$) is found
between \VC\ simulations with ideal and viscous fluid input at $\Tsw$
(dot-dashed line in Fig.~\ref{F5}a).

These observations indicate that, in Au+Au collisions at RHIC with
$(\eta/s)_\mathrm{QGP}\eq{\cal O}\left(\frac{1}{4\pi}\right)$, hadronic
dissipative effects play a larger role for the finally observed $v_2$
than QGP viscosity. This conclusion is reinforced by Fig.~\ref{F5}b with
shows (depending on centrality) 20--40\% $v_2$ suppression just from
hadronic dissipation. The \VISH\ and \VC\ comparison runs for
$\eta/s\eq1/(4\pi)$ show relatively less suppression of $v_2$ by
hadronic dissipation in \U\ since the non-zero $\eta/s$ already
suppresses hadronic $v_2$ generation in the purely hydrodynamic
run.

Obviously, the hadronic rescattering stage described by \U\ is much more
dissipative than both an ideal ($\eta/s\eq0$) and a ``perfect'' (i.e.
minimally viscous, $\frac{\eta}{s}\eq\frac{1}{4\pi}$) fluid. This raises
the question whether we could perhaps simulate the hadronic rescattering
cascade hydrodynamically by making the fluid more viscous on the hadronic
side of the quark-hadron phase tansition. The answer to this question is
explored in the following section.

%#############################################################################
\section{\VC\ with lower switching temperatures and a temperature dependent
         hadronic $\bm{(\eta/s)(T)}$}
\label{sec4}
%#############################################################################

Hydro+cascade hybrid approaches that use in their hydrodynamic modules an
equation of state that assumes chemical equilibrium in the hadron resonance
gas have little choice where to switch from the macroscopic fluid dynamic to
the microscopic Boltzmann picture: If they want to correctly reproduce
the experimentally measured final hadron yields which reflect chemical
freeze-out at $\Tchem{\,\simeq\,}165$\,MeV, the switching must be done at
that temperature. With the chemically frozen EOS s95p-PCE we can also select
lower switching temperatures and, in doing so, explore the existence of
a ``switching window'' within which, using appropriately adjusted transport
coefficients for the hydrodynamic evolution, both macroscopic and microscopic
descriptions can be used interchangeably, without affecting the final outcome.

To judge the validity of using hydrodynamics to emulate microscopic \U\
dynamics we check the final pion and proton transverse momentum spectra
and their differential elliptic flow $v_2(p_T)$. We do so for a fixed
impact parameter $b\eq7$\,fm which for the Au+Au collision system is known
to also provide a fair representation of the spectra and elliptic flow
from minimum bias collisions.

%============================ Fig. 6 ==================================
\begin{figure}[b]
\vspace{-3mm}
\includegraphics[width=0.75\linewidth,clip=,angle=270]{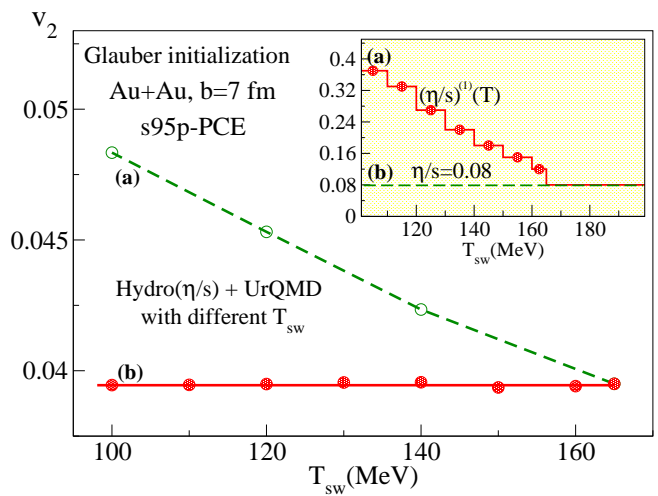}
\caption{\label{F6} (Color online) Integrated $v_2$ for pions for $b\eq7$\,fm
Au+Au, calculated from \VC\ with different switching temperatures $\Tsw$.
Case (a) assumes constant $\eta/s\eq0.08$ in \VISH, whereas case (b) uses the
temperature dependent $(\eta/s)^{(1)}(T)$ shown in the inset, which was
extracted from the integrated pion $v_2$ as described in the text.
}
\end{figure}
%======================================================================

%============================ Fig. 7 ==================================
\begin{figure*}[t]
\centering
\includegraphics[width=0.37\linewidth,clip=,angle=270]{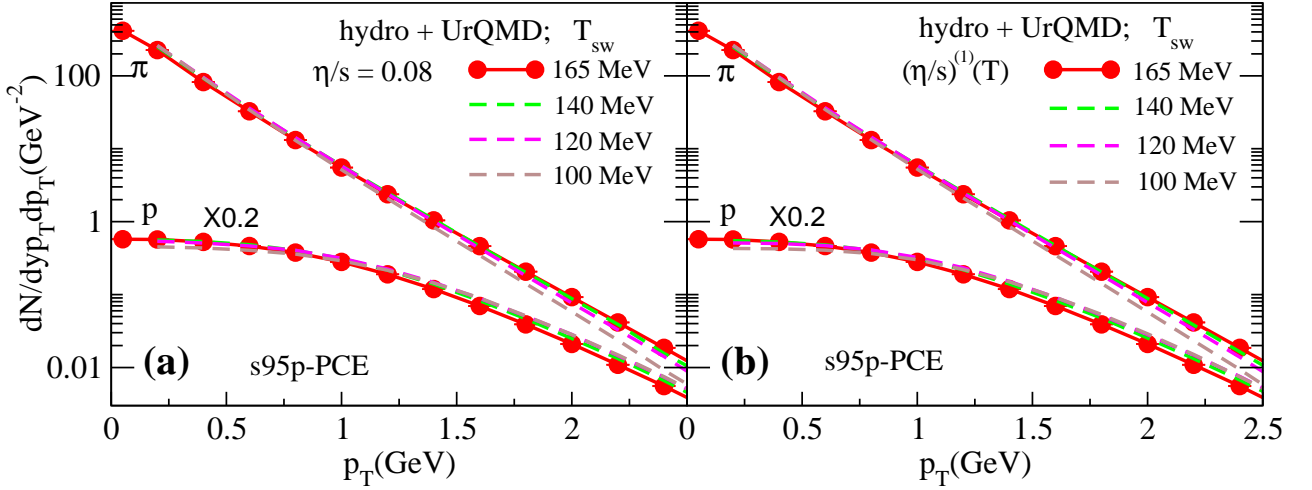}
\caption{\label{F7} (Color online) Pion and proton $p_T$-spectra from
$b\eq7$\,fm Au+Au collisions, calculated with \VC\ using different
switching temperatures $\Tsw$ and constant ($\eta/s\eq0.08$, panel (a))
or temperature dependent ($(\eta/s)^{(1)}(T)$ from Fig.~\ref{F6}, panel (b))
shear viscosities in the hydrodynamic stage.
}
\end{figure*}
%======================================================================

The most sensitive and robust \cite{Song:2010mg} observable that will
dictate our (temperature dependent) choice of $\eta/s$ in the hadronic
phase is the integrated elliptic flow $v_2$. Strictly speaking, it is
controlled by both the shear viscosity and the associated microscopic
relaxation time $\tau_\pi$ that controls the speed with which the shear
pressure tensor $\pi^{\mu\nu}$ approaches its Navier-Stokes limit
\cite{Song:2007fn}. In kinetic transport theory, both $\eta$ and
$\tau_\pi$ involve integrals over the same collision kernel
and are typically proportional to each other \cite{Moore:2008ws}.
In the QGP, where $\eta/s$ is small, $\tau_\pi$ is therefore short
of order 0.2 \fm, leading to rapid memory loss and insensitivity of the
developing elliptic flow to initial conditions for the viscous pressure
\cite{Song:2007fn}. If, however, $\eta/s$ becomes large in the hadronic
phase \cite{Csernai:2006zz,Chen:2007jq,Kapusta:2008vb,Demir:2008tr},
one should expect the corresponding relaxation time to grow similarly,
with a constant of proportionality between the dimensionless combinations
$\eta/s$ and $T\tau_\pi$ that may be different and possibly larger in
the hadronic than in the QGP phase.

Unfortunately, with our present code setup an independent determination
of the temperature dependence of $\eta/s$ and $T\tau_\pi$ in the hadronic
stage (from a comparison of \VC\ simulations with different $\Tsw$ values)
is too time consuming and thus not practical. We have therefore focused our
attention on the extraction of $(\eta/s)(T)$, holding the relation
$T\tau_\pi\eq3\,\eta/s$ fixed at all temperatures.

%%%%%%%%%%%%%%%%%%%%%%%%%%%%%%%%%%%%%%%%%%%%%%%%%%%%%%%%%%%%%%%%%%%%%%%%%
\subsection{Effective $\bm{(\eta/s)(T)}$ from \U\ with \VISH\ input}
\label{sec4a}
%%%%%%%%%%%%%%%%%%%%%%%%%%%%%%%%%%%%%%%%%%%%%%%%%%%%%%%%%%%%%%%%%%%%%%%%%

The green dashed line in Fig.~\ref{F6} shows that, if we couple \VISH\ with
constant specific entropy $\eta/s\eq0.08$ to \U\ at lower and lower switching
temperatures, we obtain larger and larger values for the total pion $v_2$.
Obviously, the hydrodynamic evolution between $\Tchem\eq165$\,MeV and
$\Tsw{\,<\,}\Tchem$ generates more additional elliptic flow than propagation
of the hadrons with \U\ during the same temperature interval. This
suggests that \U\ in the temperature region $\Tsw{\,<\,}T{\,<\,}\Tchem$
has a larger effective shear viscosity $\eta/s$ than 0.08. By lowering the
switching temperature from the starting value $\Tsw\eq165$\,MeV in small
steps $\Delta T$ and adjusting, step by step, $\eta/s$ in the interval
[165\,MeV$-n\Delta T$, 165\,MeV$-(n{-}1)\Delta T$] such that no additional
pion $v_2$ is generated when we replace \U\ evolution in this temperature
interval by \VISH\ with this adjusted $\eta/s$ value, we arrive at the
effective temperature dependent $(\eta/s)^{(1)}(T)$ shown in the inset
of Fig.~\ref{F6}) that ensures a constant pion elliptic flow $v_2$ that
is independent of the switching temperature (flat horizontal line
in Fig.~\ref{F6}). Note that this is a time consuming and labor-intensive
iterative procedure that cannot be short-circuited since the effective
$\eta/s$ extracted for the $n^\mathrm{th}$ interval turns out to depend
on the previously determined effective $(\eta/s)^{(1)}(T)$ values for
smaller $n$ values, i.e. at higher temperatures.

%%%%%%%%%%%%%%%%%%%%%%%%%%%%%%%%%%%%%%%%%%%%%%%%%%%%%%%%%%%%%%%%%%%%%%%%%%%
\subsection{Testing the effective $\bm{(\eta/s)^{(1)}(T)}$ from \U}
\label{sec4b}
%%%%%%%%%%%%%%%%%%%%%%%%%%%%%%%%%%%%%%%%%%%%%%%%%%%%%%%%%%%%%%%%%%%%%%%%%%%
%\vspace*{-4mm}

Starting at the QGP input value of $\eta/s\eq0.08$, the thus extracted
effective $(\eta/s)^{(1)}(T)$ increases by almost a factor 5 between
$T\eq\Tchem\eq165$\,MeV and $T\eq100$\,MeV, growing roughly linearly
with decreasing temperature. Figure~\ref{F7} shows that this has very little
effect on the pion and proton $p_T$-distributions: Whether one uses constant
$\eta/s\eq0.08$ (panel (a)) or the temperature dependent $(\eta/s)^{(1)}(T)$
from Fig.~\ref{F6} (panel (b)), the transverse momentum spectra for both
pions and protons exhibit very little dependence on the switching temperature
$\Tsw$. Just as for the pure hydrodynamic calculations shown in Fig.~\ref{F1},
lower switching temperatures lead to slightly flatter spectra for protons
(where the hadronic buildup of additional radial flow overwhelms the
cooling effect) and to slightly steeper spectra for pions (for which the
cooling effect dominates). Very little of the difference in the final
spectra that can be attributed to the different hadronic viscosities used
in the two panels of Fig.~\ref{F4}.

%
%============================ Fig. 8 ==================================
\begin{figure}[thb]
\includegraphics[width=0.70\linewidth,clip=,angle=270]{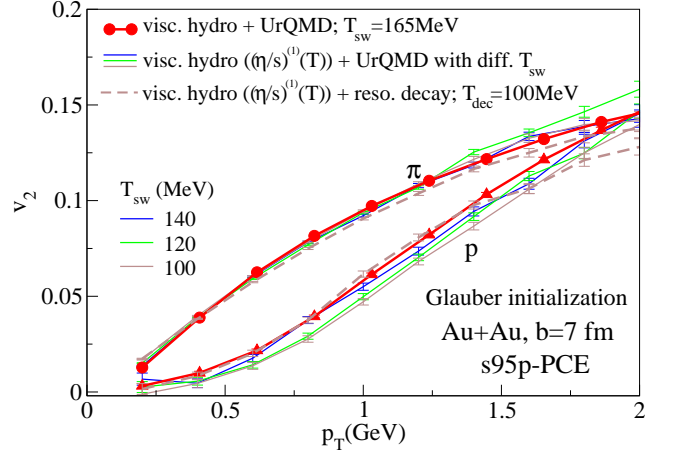}
\caption{\label{F8} (Color online) Differential $v_2(p_T)$ for pions and
protons in $b\eq7$\,fm Au+Au collisions, calculated by matching viscous
hydrodynamics with temperature dependent $(\eta/s)^{(1)}(T)$ (see
Fig.~\ref{F6}) to \U\ at different switching temperatures $\Tsw$.
}
\end{figure}
%======================================================================
%

For the elliptic flow the temperature dependence matters, as Figure~\ref{F6}
clearly demonstrates. However, Figure~\ref{F8} shows that, when using the
``correct'' temperature dependent $(\eta/s)^{(1)}(T)$ in the \VISH\ module,
\VC\ produces not only $\Tsw$-independent {\em integrated} pion elliptic
flow (by construction), but also approximately $\Tsw$-independent
{\em $p_T$-differential} elliptic flow $v_2(p_T)$ for both pions and
protons. Still, when looking at the details one observes a difference
between the solid and dashed brown lines for protons which shows that
decoupling \VISH\ at $\Tdec\eq100$\,MeV into non-interacting particles
is {\em not} the same as switching from \VISH\ to \U\ at $\Tsw\eq100$\,MeV
and letting \U\ do the kinetic freeze-out: there obviously is some
rescattering of protons in \U\ occurring at temperatures below 100\,MeV,
caused by ``pion wind'' \cite{Hung:1997du}, that moves the protons to
larger $p_T$ and depletes the proton elliptic flow at low $p_T$. Since
the solid brown line, corresponding to hydrodynamic evolution down to
100\,MeV followed by \U\ freeze-out, is different from the solid red
line with triangles, which corresponds to a switch from hydrodynamics to
\U\ already at 165\,MeV, we see that even with the ``correct'' temperature
dependent $(\eta/s)^{(1)}(T)$ hydrodynamic evolution with \VISH\ differs
somewhat from \U. However, the difference is small, and for pions its
consequences are almost negligible.

\begin{figure}[t]
\includegraphics[width=0.75\linewidth,clip=,angle=270]{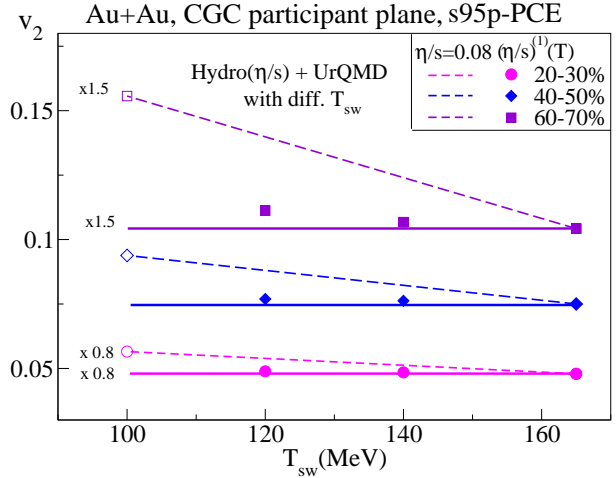}
\caption{\label{F9} (Color online) Pion $v_2$ for Au+Au collisions of
different centralities (bottom set: 20-30\%; middle set: 40-50\%;
top set: 60-70\%), calculated with \VC\ using constant $\eta/s\eq0.08$
(dashed lines) or temperature dependent $(\eta/s)^{(1)}(T)$ (solid lines)
in the fluid dynamic stage. In contrast to Figs.~\ref{F6} and \ref{F7},
we here use averaged fluctuating {\tt fKLN} initial conditions.}
\end{figure}

\begin{figure}[thb]
\includegraphics[width=0.73\linewidth,clip=,angle=270]{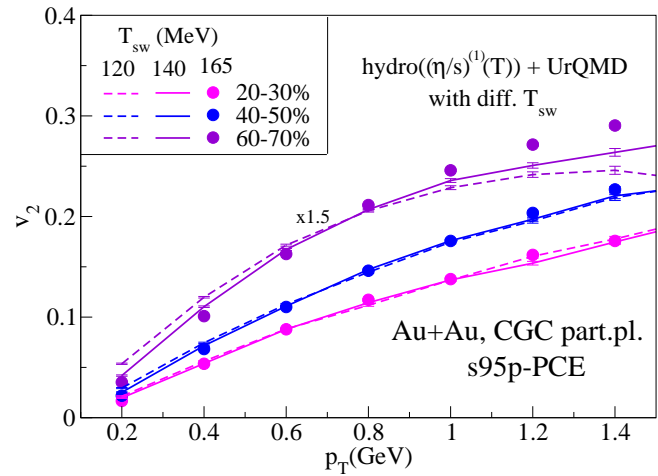}
\caption{\label{F10} (Color online) Differential $v_2(p_T)$ for pions.
Same parameters and color coding as in Fig.~\ref{F9}. }
\end{figure}

Figures~\ref{F9} and \ref{F10} show that the same $(\eta/s)^{(1)}(T)$ extracted
in Fig.~\ref{F6} from simulations with optical Glauber model initial
conditions also works for CGC motivated initial conditions that are
obtained by averaging over many fluctuating initial entropy density
profiles (see Sec.~\ref{sec2a}) and thus account for the large effects
from event-by-event fluctuations on the average initial source eccentricity
in very central and very peripheral events. For all three centrality classes
shown in these two figures we see that using the temperature dependent
$(\eta/s)^{(1)}(T)$ extracted from Fig.~\ref{F6} for the hydrodynamic evolution
%
%============================ Fig. 9 ==================================
%======================================================================
%
yields values for the integrated pion elliptic flow $v_2$ (Fig.~\ref{F9})
and shapes for the $p_t$-differential pion elliptic flow $v_2(p_T)$
(Fig.~\ref{F10}) that are independent of the switching temperature
$\Tsw$ down to $\Tsw\eq120$\,MeV (we did not probe any lower). This is
not trivial since the initial source eccentricities (and thus the produced
elliptic flows) for the centralities shown here are 20--50\% larger than
the optical Glauber ones.

%
%============================ Fig. 10 ==================================
%======================================================================
%

In the most peripheral bin (60-70\%) things begin to break down around
$\Tsw\eq120$\,MeV. The increase of integrated $v_2$ and of $v_2(p_T)$
at low $p_T$ seen in Figs.~\ref{F9} and \ref{F10} for $\Tsw\eq120$\,MeV
indicates stronger dissipative effects in \U\ at low $T$ than captured
by effective $(\eta/s)^{(1)}(T)$ from Fig.~\ref{F6}. In peripheral
collisions (i.e. for small fireballs) \VISH\ apparently can no longer
accurately mimic the microscopic kinetic evolution of \U\ at low
temperatures $\Tsw{\,\sim\,}120$\,MeV and should therefore not be used
under such conditions.

%\Blue{The
%stronger bending of $v_2(p_T)$ seen in Fig.~\ref{F10} for $\Tsw\eq120$\,MeV
%in the 60--70\% centrality bin, on the other hand, suggests that with the
%given $(\eta/s)^{(1)}(T)$ \VISH\ is already producing a $\delta f$
%correction that is too large}
%These two inferences contradict each other,
%so the correct conclusion should be that in peripheral collisions (i.e.
%for small fireballs) at low temperatures $\Tsw{\,\sim\,}120$\,MeV \VISH\
%can no longer accurately mimic the microscopic kinetic evolution of \U\
%and should therefore not be used under such conditions.

We can summarize the findings of this subsection in the following statement:
By using the temperature-dependent $(\eta/s)^{(1)}(T)$ extracted in
Sec.~\ref{sec4a} from the integrated pion elliptic flow by comparing
microscopic and macroscopic simulations of the hadron gas phase below
$\Tchem$, the macroscopic evolution code \VISH\ provides a fair description
of the microscopic \U\ dynamics down to temperatures around 120\,MeV. When
used in hybrid mode together with \U\ (as implemented in \VC), it yields
transverse momentum distributions of pions and protons (including their
elliptic flow $v_2(p_T)$) that are, to good approximation, independent of
the switching temperature in the window
120\,MeV${\,\leq\,}\Tsw{\,\leq\,}165$\,MeV.
This raises, however, two questions:
\begin{itemize}
\item
Is $(\eta/s)^{(1)}(T)$ a genuine medium property of the hadronic matter
described by \U? If it is, it should be independent of the viscous
hydrodynamic input into \U. So far we have only shown results where the
\U\ input was obtained from hydrodynamic simulations of the earlier QGP
evolution with a single viscosity value $(\eta/s)_\mathrm{QGP}\eq0.08$.
If the QGP has a larger or smaller viscosity, will \U\ continue to
propagate the correspondingly modified hydrodynamic input at $\Tsw$ as
if it were a viscous fluid with temperature dependent $(\eta/s)^{(1)}(T)$?
\item
The $(\eta/s)^{(1)}(T)$ values shown in the inset of Fig.~\ref{F6} are
much smaller (especially in the region just below $\Tchem{\,\approx\,}\Tc$)
than those computed in Ref.~\cite{Demir:2008tr} using the Kubo
formula. The Kubo formula evaluates a thermal equilibrium ensemble
expectation value of a spectral density operator, whereas the procedure
of Sec.~\ref{sec4a} extracts the transport coefficient from a comparison
of the dynamical output produced by two different evolution models in a
rapidly expanding expanding medium. What is the reason for these different
shear viscosities?
\end{itemize}
In the following subsection we describe a study that yields at least partial
answers to these questions.

%%%%%%%%%%%%%%%%%%%%%%%%%%%%%%%%%%%%%%%%%%%%%%%%%%%%%%%%%%%%%%%%%%%%%%%%%%%%%
\subsection{Non-universality of the effective \U\ viscosity
            $\bm{(\eta/s)^{(1)}(T)}$}
\label{sec4c}
%%%%%%%%%%%%%%%%%%%%%%%%%%%%%%%%%%%%%%%%%%%%%%%%%%%%%%%%%%%%%%%%%%%%%%%%%%%%%

\begin{figure}[b]
\includegraphics[width=0.70\linewidth,clip=,angle=270]{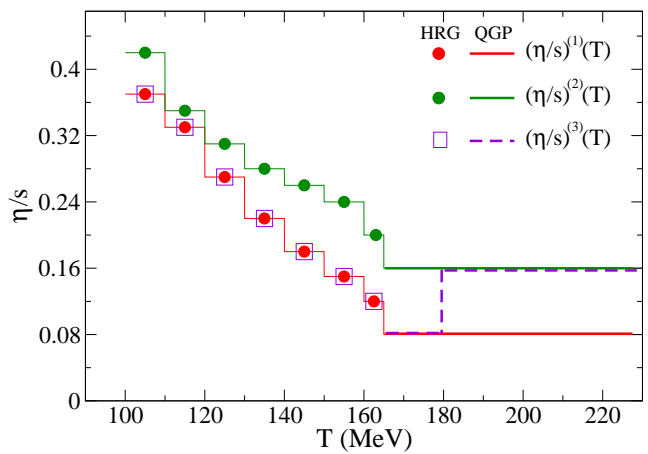}
\caption{\label{F11} (Color online) Temperature dependent effective
$(\eta/s)(T)$ extracted as in Fig.~\ref{F6} from viscous hydrodynamic
emulations of \U\ dynamics, for three different viscous fluid initializations
of the \U\ stage (see text for discussion).
}
\end{figure}

To address the first question we redid the analysis of Sec.~\ref{sec4a}
using a twice larger QGP viscosity, $(\eta/s)_\mathrm{QGP}\eq0.16$.
This means that the \U\ module of \VC\ is initialized with somewhat
different density and flow profiles and, in particular, a shear viscous
pressure tensor $\pi^{\mu\nu}{\,\approx\,}2\eta\sigma^{\mu\nu}$ that is
roughly twice as large as before when we used
$(\eta/s)_\mathrm{QGP}\eq0.08$.\footnote{Here $\sigma^{\mu\nu}\eq\nabla^{\left
  \langle\mu\right.}u^{\left.\nu\right\rangle}$ is the velocity shear tensor,
  see \cite{Song:2007fn} for details. Since $\eta/s$ is small in the QGP,
  the associated relaxation time $\tau_\pi$ is much shorter than the inverse
  of the local expansion rate on the switching surface, hence $\pi^{\mu\nu}$
  does not stray far from its Navier-Stokes value
  $\pi^{\mu\nu}\eq2\eta\sigma^{\mu\nu}$ on that surface.}
Going through the same iterative procedure as before of adjusting the
specific shear viscosity $\eta/s$ temperature interval by temperature
interval such that it generates in \VISH\ exactly the same amount of
total pion $v_2$ as \U, we obtain the upper set of solid green dots in
Fig.~\ref{F11}, labeled as $(\eta/s)^{(2)}(T)$. It is obviously different
from and larger than $(\eta/s)^{(1)}(T)$. In particular, at the highest
switching temperature $\Tsw\eq165$\,MeV, $(\eta/s)^{(2)}(T)$ starts out
close to the value of 0.16 used in the evolution of the QGP stage,
whereas $(\eta/s)^{(1)}(T)$ starts out almost half as small, with
a magnitude close to the QGP value of 0.08 used in the first extraction.

%============================ Fig. 11 ==================================
%======================================================================

The ``effective \U\ viscosity'' $(\eta/s)(T)$ extracted by this procedure
apparently ``remembers'' the prior QGP history of the fireball and its
transport properties. This means that it does not describe an intrinsic
medium property of the hadron resonance gas in \U. Further insight
into this puzzle is gained by looking at the purple dashed line and
open squares in Fig.~\ref{F11}, labeled by $(\eta/s)^{(3)}(T)$: Here we
initialize \U\ with viscous hydrodynamic output from a run with
$(\eta/s)_\mathrm{QGP}\eq0.16$, but assume that just before hadronization,
at a temperature of 180\,MeV, the QGP viscosity suddenly drops to
half that value, $(\eta/s)_\mathrm{QGP}\eq0.08$. In this case, the
extracted ``effective \U\ viscosity'' $(\eta/s)^{(3)}(T)$ comes out
identical to $(\eta/s)^{(1)}(T)$, i.e. as if the QGP phase had evolved
with minimal shear viscosity throughout its life, and not just at the
very end of its history after its temperature dropped below 180\,MeV.

Evolving the QGP medium from identical initial conditions with a twice
larger $\eta/s$ value of 0.16 leads to noticeably different density and
flow profiles on the $\Tsw\eq165$\,MeV switching surface compared to
those for $\eta/s\eq0.08$. These differences cannot be undone by changing
$\eta/s$ back to 0.08 within 15\,MeV of the critical temperature. The
equality of $(\eta/s)^{(1)}(T)$ and $(\eta/s)^{(3)}(T)$ shows that they
are not controlled by the energy density and flow velocity profiles on
the switching surface (which are different for the solid red and dashed
purple lines), but instead by the shear pressure tensor. Due to the short
relaxation times $\tau_\pi$ in the QGP phase, we can approximate
$\pi^{\mu\nu}{\,\approx\,}2\eta\sigma^{\mu\nu}$, so a factor of 2 difference
in $\eta/s$ leads to roughly a factor 2 difference in $\pi^{\mu\nu}$ on
the switching surface. (The differences in the entropy density and
velocity shear tensor, although noticeable, are not anywhere close
to a factor 2.) At this level of precision, the hydrodynamic output
corresponding to the green solid line (from a QGP evolving with
$(\eta/s)_\mathrm{QGP}\eq0.16$ all the way to $\Tsw\eq165$\,MeV) features
a viscous pressure $\pi^{\mu\nu}$ that is twice as big as that for
the red solid line ($(\eta/s)_\mathrm{QGP}\eq0.08$), whereas the one
corresponding to the purple dashed line is roughly equal to that of the
red solid line. The ``effective \U\ viscosities'' $(\eta/s)^{(n)}(T)$
seem to reflect and remember these relationships.

We can understand this by recalling the kinetic theory relation
$\pi^{\mu\nu}\eq\int\frac{d^3p}{E} p^{\left\langle\mu\right.}
p^{\left.\nu\right\rangle}\,\delta f$ between the viscous pressure tensor
and the deviation $\delta f$ of the phase-space distribution function
from local thermal equilibrium. The value of $\pi^{\mu\nu}$ on the switching
surface controls the magnitude of the non-equilibrium contribution
$\delta f \sim p^\mu p^\nu \pi_{\mu \nu}$ to the hadron momentum
distributions sampled by the hydro-to-micro converter \con\ which
generates the \U\ input. Apparently, collisions and other interactions
in \U\ are too infrequent or inefficient in transferring momenta to
quickly relax these non-equilibrium distributions. As long as the
deviations $\delta f$ persist in a form close to their initial values,
they contribute a term $\pi^{\mu\nu}$ to the \U\ energy-momentum tensor
that, if one writes it in hydrodynamic language as $\pi^{\mu\nu}\eq2
\eta\sigma^{\mu\nu}$, requires a value of $\eta/s$ that is close to the
value used in the generation of the \U\ input with \con. As \U\ evolves
the momentum distributions, $\pi^{\mu\nu}$ evolves, too, and when we
fit the \VISH\ transport properties to those of the \U\ cascade assuming
short relaxation times, we end up fitting this evolving $\pi^{\mu\nu}$
instead of extracting a genuine transport coefficient for \U.

The $(\eta/s)^{(n)}(T)$ curves shown in Fig.~\ref{F11} exhibit a tendency
to approach each other at lower temperatures. This may suggest a loss
of memory of the initial viscous pressure components on a time scale
comparable to the cooling time needed to cool the system from
$\Tchem\eq165$\,MeV to somewhere around 100\,MeV. This time is longer than
the one assumed in our viscous hydrodynamic simulations of the \U\ stage,
$T\tau_\pi\eq3\eta/s$. Indeed, there are other indications \cite{notes}
that \U\ may have a much longer relaxation time than given by the classical
kinetic theory result for massless Boltzmann particles,
$\tau_\pi\eq6\eta/(sT)$ \cite{Israel:1976tn}. For the time being, further
exploration of this issue will have to wait.

%%%%%%%%%%%%%%%%%%%%%%%%%%%%%%%%%%%%%%%%%%%%%%%%%%%%%%%%%%%%%%%%%%%%%%%%%%%%%
\section{Discussion and conclusions}
\label{sec5}
%%%%%%%%%%%%%%%%%%%%%%%%%%%%%%%%%%%%%%%%%%%%%%%%%%%%%%%%%%%%%%%%%%%%%%%%%%%%%

In this article we presented \VC, a new hybrid approach for the bulk
evolution of viscous QCD matter created in ultrarelativistic heavy-ion
collisions that combines a macroscopic viscous hydrodynamic description
of the early (dense) QGP phase with a microscopic Boltzmann cascade for the
late (dilute) hadron resonance gas stage. The model merges the economy of
a macroscopic description for the first $4{-}10$\,fm/$c$ (depending on
centrality) after thermalization, when the matter is close to local thermal
equilibrium, with the precision of a microscopic approach for the late
hadronic stage when, after hadronization of the QGP, the mean free paths
for the hadronic medium constituents increase rapidly, the matter moves
farther and farther away from local equilibrium, and finally decouples
into non-interacting, free-streaming particles.

The hydrodynamic and Boltzmann cascade components of \VC\ are connected
via the hydro-to-micro converter \con, a Monte Carlo event generator that
samples the Cooper-Frye phase-space distributions from the viscous
hydrodynamic output along a switching surface of constant temperature
$\Tsw$ and injects the resulting particles into \U\ for further
microscopic propagation.

By using a microscopic cascade approach for the late hadronic stage in
heavy-ion collisions, the complex microscopic dynamics of chemical and
kinetic freeze-out is accounted for without extraneous parameters, removing
a major source of uncertainty in connecting final hadron spectra with
transport properties of the initial QGP phase. With the help of \VC\
a reliable extraction of the QGP shear viscosity $(\eta/s)_\mathrm{QGP}$
from experimental RHIC data, with good theoretical control of the
associated uncertainties, has now become possible \cite{Song:2010mg}.

In default mode, \VC\ is used with switching temperature
$\Tsw\eq\Tchem\eq165$\,MeV. Since the experimentally determined chemical
freeze-out temperature $\Tchem{\,\approx\,}165$\,MeV
\cite{BraunMunzinger:2001ip} approximately agrees with the best presently
available theoretical estimate for the (pseudo)critical temperature for
the chiral phase transition in QCD \cite{Bazavov:2009zn,Borsanyi:2010zh},
this is the highest temperature for which a microscopic description in
terms of colliding hadrons with vacuum masses and decay widths makes
sense, i.e. the highest temperature for which \U\ can be reliably used.
In default mode the system is therefore described by a chemical equilibrium
EOS in the hydrodynamic stage, followed by a microscopic stage in which
first chemical and then kinetic equilibrium are broken automatically by
the microscopic scattering dynamics. In this mode no chemically frozen
EOS is needed.

The main thrust of the present work has been to investigate whether
\VC\ can be used with lower switching temperatures. Since the microscopic
evolution of the hadron resonance gas via \U\ is numerically much more
costly than a macroscopic hydrodynamic description, one would like to
use hydrodynamics as long as possible and switch to \U\ only when the
macroscopic approach is no longer reliable. We showed that in that case
using a chemically frozen hadron resonance gas EOS in the hydrodynamic
evolution below $\Tchem$ is compulsory. Without such a realistic
EOS, which takes into account that at temperatures below $\Tchem$ the stable
particle yields must not change anymore, one cannot find a single value
for the kinetic decoupling temperature in \VISH\ that simultaneously
reproduces the microscopically evolved pion and proton transverse momentum
spectra from \VC. For the differential elliptic flow $v_2(p_T)$ using
a chemical equilibrium hadronic EOS can lead to quite misleading
results.

Using the chemically frozen EOS s95p-PCE \cite{Huovinen:2009yb,Shen:2010uy}
one finds that, with the judicious choice $\Tdec\eq100$\,MeV for the kinetic
decoupling temperature, the pion and proton spectra from \VC\ can be
reasonably well reproduced by a purely hydrodynamic calculation with
\VISH, without even a need for changing the specific shear viscosity
$\eta/s$ between the QGP and HRG stages. The same does not hold, however,
for the elliptic flow: since at RHIC energies the fireball is still
out-of-plane elongated when its matter enters the HRG phase, the
momentum anisotropy continues to grow during the hadronic stage, and
\U\ is less efficient in converting the residual source eccentricity
into elliptic flow than \VISH\ if one uses the same minimal $\eta/s$ in
both phases. Since we also showed that, under RHIC conditions, the total
suppression of elliptic flow below its ideal fluid limit is dominated
by the hadronic stage (Fig.~\ref{F5}), this demonstrates unequivocally
that, if one wants to avoid or minimize the cost of using a microscopic
approach like \U, great care must be taken to use the correct transport
properties of the hadronic phase in \VISH\ below $\Tc$.

We therefore tried to improve the hydrodynamic description of the HRG by
increasing $\eta/s$ in the hadronic phase. To extract the temperature
dependence of $\eta/s$ in the hadronic phase as described by \U, we
lowered $\Tsw$ in small intervals and adjusted $\eta/s$ in \VISH\ in
each newly covered temperature interval such that \VISH\ produced exactly
as much additional $v_2$ for pions as \VC\ resp. \U\ did. We found a
function that we called $(\eta/s)^{(1)}(T)$ that started at
$\Tsw\eq\Tchem$ close to the value used in the QGP phase and then
increased dramatically, by almost a factor 5, as we lowered $\Tsw$ to
100\,MeV. Surprisingly, this $(\eta/s)^{(1)}(T)$ was much smaller than
the corresponding values previously extracted from \U\ using the Kubo
formula \cite{Demir:2008tr}.

With this $(\eta/s)^{(1)}(T)$, \VISH\ is able to reproduce quite well
the transverse momentum spectra as well as not only the integrated, but
also the $p_T$-differential elliptic flow of pions and protons calculated
with the full hybrid code \VC\ (Figs.~\ref{F7} and \ref{F8}). It continues
to do so if we replace the optical Glauber model initial conditions in
\VISH\ by an ensemble average of fluctuating initial conditions from the
{\tt fKLN} model (Figs.~\ref{F9} and \ref{F10}). But it fails badly
when we replace the hydrodynamic input into \U\ by one that was
calculated with a different $(\eta/s)_\mathrm{QGP}$. In fact, the effective
hadronic $(\eta/s)(T)$ extracted from \U\ by demanding that both \VISH\ and
\VC\ produce the same total pion elliptic flow appears to track the shear
viscosity of the preceding QGP phase: lower values of $(\eta/s)_\mathrm{QGP}$
produce lower values of $(\eta/s)(T)$ below $\Tc$, and {\em vice versa}.
Clearly, the extracted hadronic $(\eta/s)(T)$ is not an intrinsic medium
property of the hadron resonance gas in \U, but a parameter that preserves
some memory of the QGP transport properties (in fact, only of the
transport properties at temperatures just above $\Tchem$, see
Fig.~\ref{F11}).

Through a sequence of cross-checks we convinced ourselves that our method
to extract an effective $(\eta/s)(T)$ for the hadronic matter in \U\
failed because in \U\ the viscous corrections in the energy-momentum tensor
relax to their Navier-Stokes values much more slowly than assumed in
\VISH. As a consequence, the extracted $\eta/s$ fitted the value of
the viscous pressure tensor $\pi^{\mu\nu}$ inherited by \U\ from \VISH\
(and thus characterized by the $\eta/s$ value in the preceding QGP phase)
instead of the shear viscosity characteristic of \U\ matter.

An independent extraction of both the relaxation time $\tau_\pi$ and the
shear viscosity from \U\ is presently beyond our technical means. Our
results suggest that it may be possible to achieve a good emulation of
\U\ dynamics with viscous hydrodynamics, by using \VISH\ with larger
specific shear viscosities $\eta/s$ combined with larger relaxation times
$\tau_\pi$ in the hadronic phase than those assumed in the present work
(which were constrained by the relation $T\tau_\pi\eq3\,\eta/s$). We
are skeptic, however, about the prospects for finding a pair of functions
$(\eta/s)(T)$ and $T\tau_\pi(T)$ that work universally, {\em i.e.} that
yield good viscous hydrodynamic emulations of \U\ independent of the
initial input into \U\ (in particular, independent of the initial values
of the viscous terms in $T^{\mu\nu}$ that are to be further evolved with
\U). The reason for our skepticism is that large relaxation times $\tau_\pi$
indicate a fundamental breakdown of the viscous hydrodynamic framework:
They appear as second-order corrections to ideal fluid dynamics in a
systematic gradient expansion, and if they lead to large excursions of
the viscous pressure $\pi^{\mu\nu}$ away from its first-order Navier-Stokes
value $\pi^{\mu\nu}\eq2\eta\sigma^{\mu\nu}$, this indicates that the gradient
expansion is no longer converging.

%============================ Fig. 12 =================================
\begin{figure*}[t]
\includegraphics[width=0.35\linewidth,clip=,angle=270]{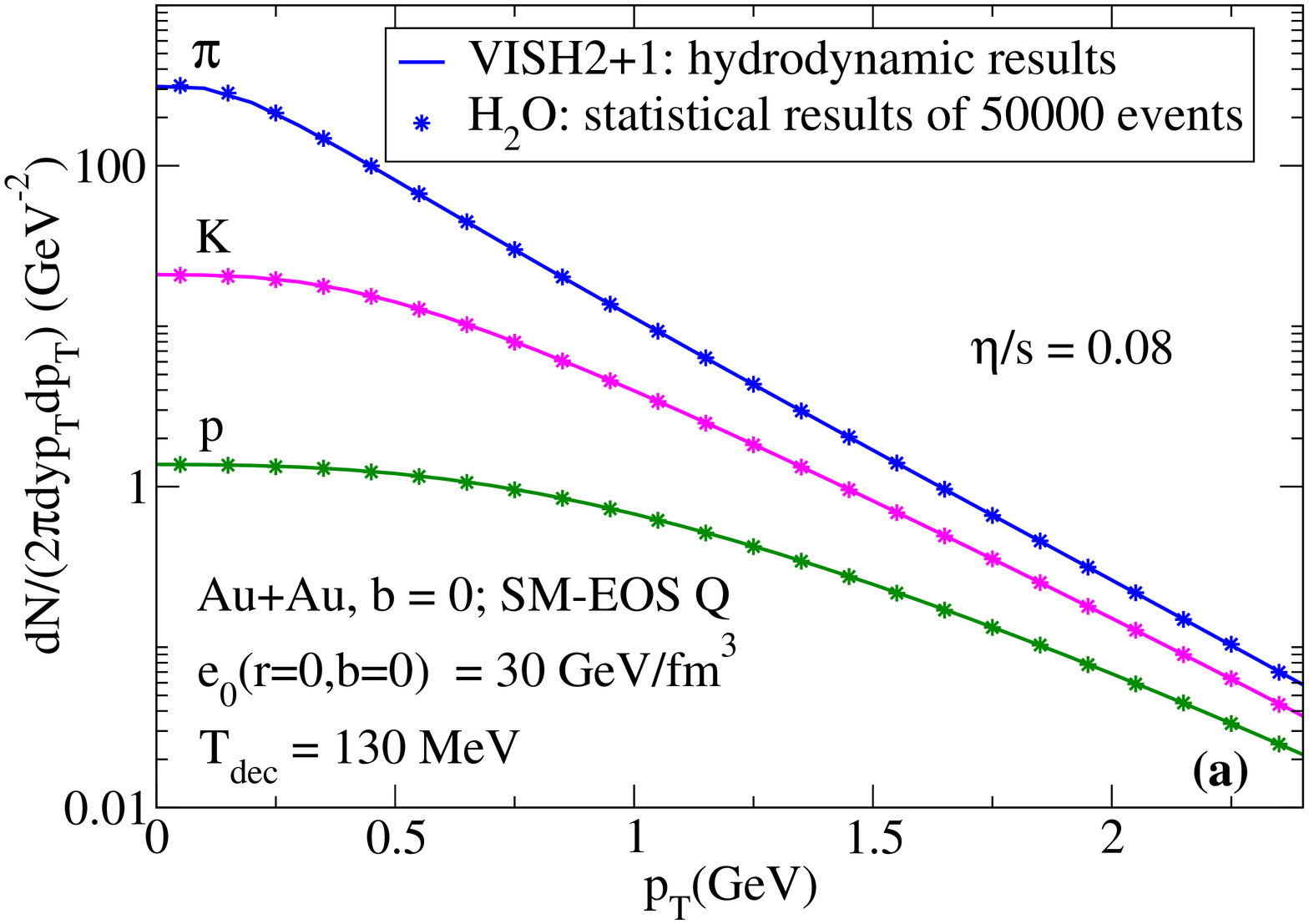}
\includegraphics[width=0.35\linewidth,clip=,angle=270]{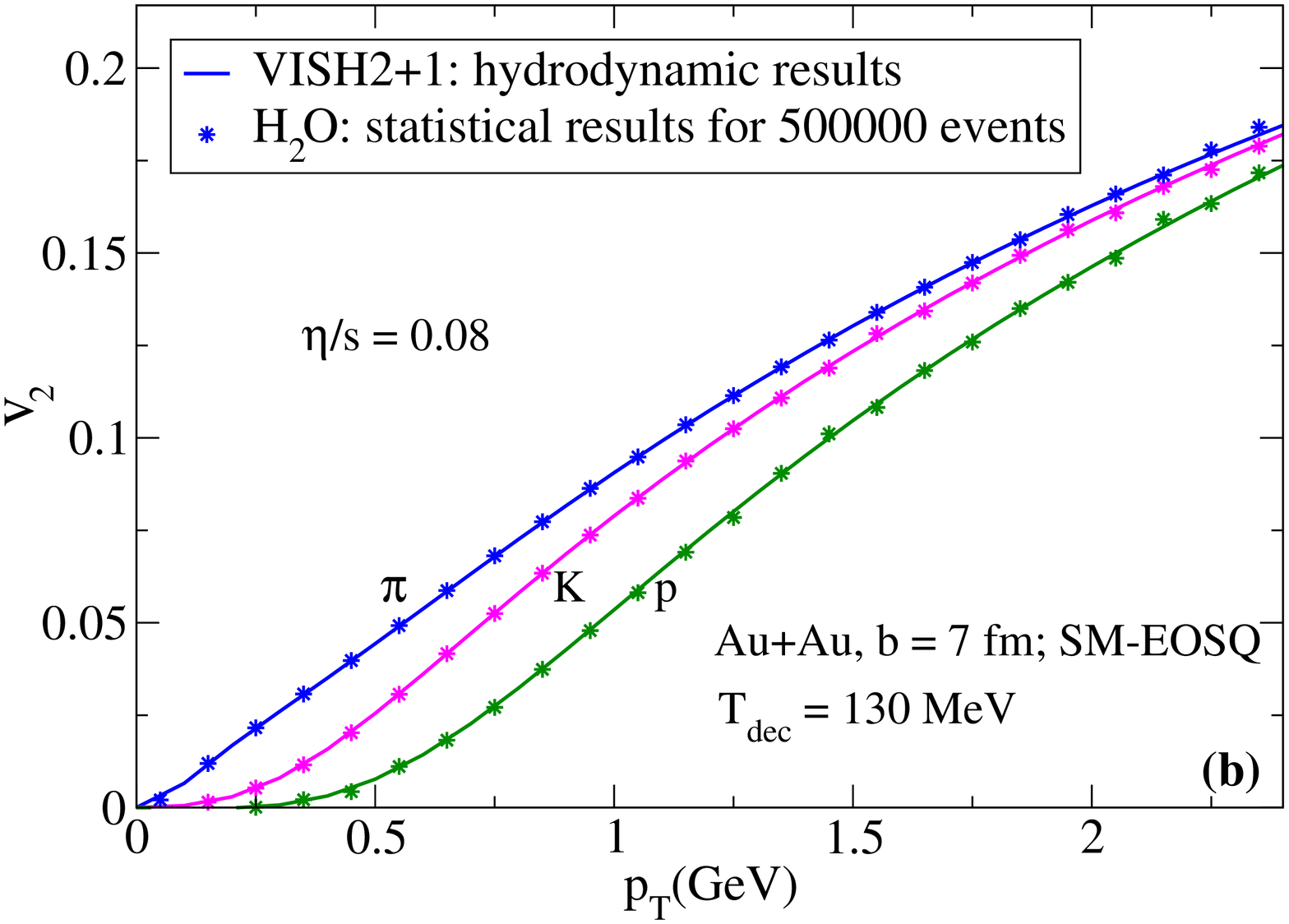}
\vspace*{-2mm}
\caption{\label{F12} (Color online) $p_T$ spectra for central Au+Au
collisions and $v_2(p_T)$ for non-central Au+Au collisions at b=7 fm, from
\VISH\ and \con.
\vspace*{-3mm}
}
\end{figure*}
%======================================================================

With this state of present knowledge we conclude that there exists no
``switching window'' of temperatures below $\Tchem$ for safely switching
from \VISH\ to \U, {\it i.e.} there is no temperature interval below
$\Tchem$ in which \VISH\ and \U\ can both be used equally well to describe
the evolution of the expanding medium created in relativistic heavy-ion
collisions. Quantitative compari\-sons with experimental data, with the goal
of extracting from measured hadron spectra precise information of QGP
transport properties, will therefore necessarily require the use of a
hybrid code like \VC\ in which the dynamics of the late hadronic stage
(at all temperatures below $\Tchem$) is described microscopically.

%%%%%%%%%%%%%%%%%%%%%%%%%%%%%%%%%%%%%%%%%%%%%%%%%%%%%%%%%%%%%%%%%%%%%%%%%%

\acknowledgments{We gratefully acknowledge fruitful discussions with V.~Koch
and T. Hirano, whom we also thank for providing the averaged initial density
profiles for fluctuating CGC initial conditions. We thank P. Huovinen for
sending us the EOS s95p-PCE before publication and T. Riley and C. Shen for
providing a fit function \cite{Shen:2010uy} for this EOS and implementing it
into {\tt VISH2+1}. This work was supported by the U.S.\ Department of Energy
under contracts DE-AC02-05CH11231, DE-FG02-05ER41367, \rm{DE-SC0004286}, and
(within the framework of the Jet Collaboration) \rm{DE-SC0004104}. We
gratefully acknowledge extensive computing resources provided to us by
the Ohio Supercomputer Center.
}

\appendix

%%%%%%%%%%%%%%%%%%%%%%%%%%%%%%%%%%%%%%%%%%%%%%%%%%%%%%%%%%%%%%%%%%%%%%%%%%%
\section{Verification of the \VISH\ to \U\ converter}
\label{appa}
%%%%%%%%%%%%%%%%%%%%%%%%%%%%%%%%%%%%%%%%%%%%%%%%%%%%%%%%%%%%%%%%%%%%%%%%%%%

In \VC\, the connection between \VISH\ and \U\ is realized by the
Monte-Carlo particle generator \con, which is briefly described in
Sec.~\ref{sec2b}. This appendix details the verification of \con\
against the analytical Cooper-Frye formula on which it is based.

\con\ randomly generates an ensemble of particles in momentum  and position
space based on the differential Cooper-Frye formula, Eq.~(\ref{Cooper}), for
each collision event, which can subsequently be used as initial configuration
for \U. Here, we directly perform a statistical analysis of sufficiently
many such initial configurations obtained from \con\ to generate smooth
particle spectra and elliptic flow curves, which we then compare to a direct
numerical integration of Eq.~(\ref{Cooper}) along the freeze-out surface
generated by \VISH.

Figure~\ref{F12} shows $p_T$-spectra for pions, protons and kaons in
central (a) and their differential $v_2(p_T)$ in non-central (b) Au+Au
collisions, obtained from \VISH\ directly and via \con, respectively. To
specifically test the ``viscous part'' of the \con\ event generator, we
select a constant specific shear viscosity $\eta/s\eq0.08$. With a
sufficiently large number of events, the statistical results from \con\
exactly reproduce the hydrodynamic spectra and $v_2$ directly obtained
from \VISH.

%============================ Fig. 13 ==================================
\begin{figure}[b]
\vspace*{-4mm}
\includegraphics[width=0.75\linewidth,clip=,angle=270]{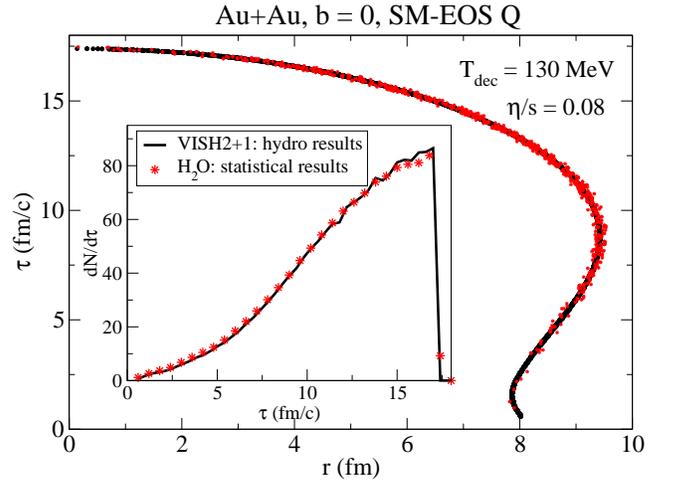}
\caption{\label{F13} (Color online) \VISH: hydrodynamic freeze-out surface
for central ($b\eq0$) Au+Au collisions with decoupling temperature
$\Tdec\eq130\ \mathrm{MeV}$. \con: pion emission points along the
freeze-out surface from a single event. Inset: pion emission rate
along the freeze-out surface.
\vspace*{-4mm}
}
\end{figure}
%======================================================================

The above hadron spectra are generated on a decoupling hypersurface with
$\Tdec\eq130$\,MeV.\footnote{We here selected the ``default''
    $\Tdec\eq130$\,MeV for EOS-Q, but the quality of the code verification
    results does not depend on the chosen decoupling/switching temperature.}
In Fig.~\ref{F13}, we show the 2-dimensional $r{-}\tau$ freeze-out
hypersurface for central Au+Au collisions. The black line is the hydrodynamic
freeze-out hypersurface obtained from \VISH, and the red symbols are the pion
emission points taken from a single \con\ event, all of which fall on the
\VISH\ freeze-out hypersurface. The small deviations which can be seen are
due to the finite position resolution in \con, which randomly samples
particle positions on the freeze-out hypersurface within $\Delta r\eq0.1$\,fm.
Fig.~\ref{F13} also shows that only few particles come from the very
early stage of the fireball evolution, while most of the particles are
emitted during the middle and late stages when the flow velocity is fully
developed. This is illustrated in the inset of Fig.~\ref{F13}, which shows
the emission rate for pions as a function of time (along the freeze-out
hypersurface). Again the statistical results from \con\ are in excellent
agreement with the hydrodynamic results from \VISH\, showing that \con\
correctly reproduces the particle emission rates of \VISH\ along the
freeze-out hypersurface.

%============================ Fig. 14 ==================================
\begin{figure}[t]
\includegraphics[width=0.7\linewidth,clip=,angle=270]{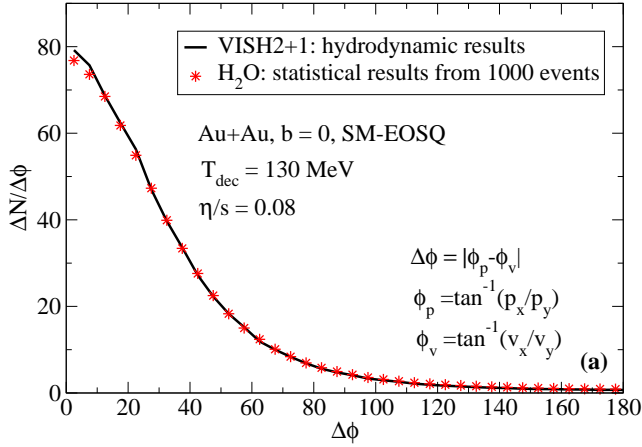}
\caption{\label{F14} (Color online) Left: Pion emission probabilities as
a function of emission angle $\Delta\phi\eq|\phi_p{-}\phi_v|$ from \VISH\
and \con.}
\end{figure}
%======================================================================

Momenta and positions of the produced particles are correlated through
Eq.~(\ref{Cooper}). For example, the equilibrium distribution $f$ for
$\mu\eq0$ can be explicitly written as
\begin{equation}
\label{A1}
  f_\mathrm{eq} = \frac{1}{e^{\gamma_\perp[m_T\cosh(y{-}\eta_s)
                         - p_T v_\perp \cos(\phi_p{-}\phi_v)]/T}\pm 1}.
\end{equation}
This shows directly the correlation between the transverse flow velocity
$\bm{v}_\perp$ and transverse momentum $\bm{p}_T$ of the particles, as well
as the correlation between momentum rapidity $y$ and space rapidity $\eta_s$.
(Here $\phi_p\eq\arctan(p_x/p_y)$ is the angle of the particle transverse
momentum and $\phi_v\eq\arctan(v_x/v_y)$ is the angle of the transverse
flow velocity.)

In fact, the above form of the distribution function leads to enhanced
particle production along the flow velocity direction, but suppresses
particle production in the opposite direction. This is shown in Fig.~\ref{F14}
which shows the particle production rate as a function of the relative
emission angle $\Delta\phi\eq|\phi_p{-}\phi_v|$. Both the hydrodynamic
result from \VISH\ as well as the statistical result from \con\ show
that $\Delta N/\Delta\phi$ reaches a peak when $\Delta\phi$ is zero
(particle emission along the flow velocity direction), then rapidly
decreases with increasing $\Delta \phi$, and reaches a minimum for
$\Delta\phi\eq180^\circ$. The excellent agreement between the \VISH\ and
\con\ results show that \con\ correctly describes the particle momentum
and flow velocity correlations encoded in Eq.~(\ref{Cooper}).

\begin{figure}[tbh]
\vspace*{-0mm}
\includegraphics[width=0.7\linewidth,clip=,angle=270]{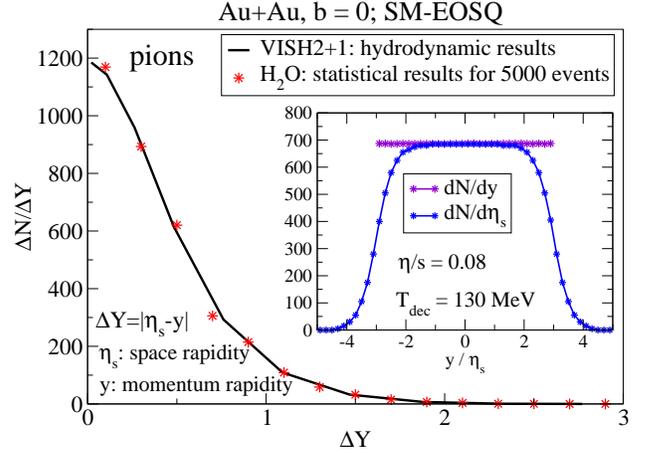}
\caption{\label{F15} (Color online) Left: Pion emission probabilities as
a function of emission rapidity $\Delta Y\eq|\eta_s{-}y|$ for \VISH\ and
\con. Inset: Pion production as a function of momentum rapidity $y$ and
space-time rapidity $\eta_s$, from \con.}
\end{figure}

The Bjorken approximation in (2+1)-dimensional viscous hydrodynamics leads
to a uniform particle density as a function of momentum-space rapidity and
likewise a uniform particle density as a function of space-time rapidity.
Eq.~(\ref{Cooper}) (together with Eq.~(\ref{A1})) shows the correlation
between momentum-space and space-time rapidity, which prevents \con\ from
generating particles with independent momentum-space and space-time
rapidities. In practice, \con\ first randomly generates the momentum-space
rapidity $y$ for each hadron within a finite pre-defined range (for example
between -3 and +3) and then samples the space-time rapidity $\eta_s$ through
Eq.~(\ref{Cooper}). The finite range of $y$, together with the $y{-}\eta_s$
correlation, leads to a decrease of $dN/d\eta_s$ in the boundary region of
$\eta_s$ as shown in the inset of Fig.~\ref{F15}.

Fig.~\ref{F15} shows the correlation between $y$ and $\eta_s$ for pions.
In excess of 90\% of pions are produced at a space-time rapidity which
lies within one unit of its momentum-space rapidity. In other words, most
of the particles with forward (backward) momentum-space rapidity are
generated in forward (backward) space-time rapidity regions. Fig.~\ref{F15}
also shows that the hydrodynamic results from \VISH\ and statistical
results from \con\ agree very well with each other, indicating that \con\
correctly describes the momentum-space and space-time rapidity correlation
in particle production.

%============================ Fig. 15 ==================================
%======================================================================

%%%%%%%%%%%%%%%%%%%%%%%%  References %%%%%%%%%%%%%%%%%%%%%%%%%%%%%%%%%%%%%%%%%

\end{document}